\documentclass[prd,reprint]{revtex4-1}

\usepackage{amsmath,amssymb,amsfonts,amsthm}
\usepackage{graphicx}
\newcommand{\comment}[1]{}

\newcommand{\Dr}{\Delta_r}
\newcommand{\Dt}{\Delta_\theta}

\newcommand{\sL}{\,\textrm{sign} L\,}

\begin{document}

\title{Observables for bound orbital motion in axially symmetric space-times}

\author{Eva Hackmann}
\email{eva.hackmann@zarm.uni-bremen.de}
\affiliation{ZARM, University of Bremen, Am Fallturm, 28359 Bremen, Germany}

\author{Claus L\"ammerzahl}
\email{claus.laemmerzahl@zarm.uni-bremen.de}
\affiliation{ZARM, University of Bremen, Am Fallturm, 28359 Bremen, Germany}
\affiliation{Institute for Physics, University Oldenburg, 26129 Oldenburg, Germany}

\begin{abstract}
The periastron shift and the Lense-Thirring effect of bound orbital motion in a general axially symmetric space-time given by Pleba\'nski and Demia\'nski are analyzed. We also define a measure for the conicity of the orbit and give analytic expressions for the observables in terms of hyperelliptic integrals and Lauricella's $F_D$ function. For an interpretation of these analytical expressions, we perform a post-Schwarzschild and a post-Newton expansion of these quantities. This clearly shows the influence of the different space-time parameters on the considered observables and allows to characterize Kerr, Taub-NUT, Schwarzschild-de Sitter, or other space-times.
\end{abstract}

\maketitle


\section{Introduction}

Axially symmetric vacuum solutions of the Einstein field equations are used to describe a wide range of black holes appearing in the universe. The most prominent of these solutions is the Kerr space-time \cite{Kerr63}, which describes an axially symmetric rotating black hole. Generalizations of these are the charged Kerr-Newman black holes, the Kerr-de Sitter space-time which incorporates a non-vanishing cosmological constant, the Kerr-Taub-NUT space-time which includes the NUT-charge, or some combination of these. All these space-times can be seen as special cases of the family of electrovac space-times of Petrov type D given by the Pleba\'nski-Demia\'nski class of solutions  \cite{PlebanskiDemianski76}. They are characterized by seven parameters: the mass, the rotation around the symmetry axis, the electric and magnetic charge, the NUT charge, the cosmological constant, and the acceleration of the gravitating source.

Shortly after the introduction of general relativity, Einstein found that for every revolution of a planetary orbit the point of the shortest distance between particle and central object, the periapsis, is shifted in the direction of rotation of the test particle  \cite{Einstein1915}. Indeed, the explanation of the anomalous shift of Mercury's perihelion, together with the observation of light deflection, constituted the breakthrough of general relativity. Already in 1918, Lense and Thirring found that bound orbital motion around an axially symmetric rotating black hole is perturbed not only compared to the Newtonian case but also to the non-rotating case \cite{LenseThirring18}. The rotation of the central object causes an additional shift of the periapsis and a precession of the orbital plane. For the more exotic Taub-NUT space-time Misner and Taub \cite{MisnerTaub69} showed that the geodesic motion takes place on a cone which, if slit open and flattened, defines the orbital plane as it would be for vanishing NUT charge. These three examples show that the parameters of the space-time affect the observables of bound orbital motion and that in turn these observables may be used to characterize a space-time. In this paper we will investigate these orbital effects starting from the general Pleba\'nski-Demia\'nski space-time with vanishing acceleration of the gravitating source.

The only way to get access to these characteristics of black holes is through the orbits of particles and light around these black holes. They are given as solutions of the geodesic equation describing the motion of test particles and light rays in a given space-time. For a thorough investigation of the physical properties of orbits analytical solutions of the geodesic equation are most useful. Hagihara \cite{Hagihara31} was the first to find an analytical solution of the geodesic equation in Schwarzschild space-times using elliptic functions. He also classified all possible types of orbits in terms of the energy and the angular momentum of the test particle. The complete set of solutions for charged particle motion in Reissner-Nordstr\"om space-times have been presented only recently by Grunau and Kagramanova \cite{GrunauKagramanova11}. Shortly after the discovery of the Kerr solution in 1963 \cite{Kerr63}, a number of authors studied the geodesic motion in this space-time. Their results were reviewed and extended by Chandrasekhar \cite{Chandrasekhar83}. A treatise of the geodesics in Kerr space-time of the same completeness as \cite{Hagihara31} was given only recently by Slez\'{a}kov\'{a} \cite{Slezakova06} and for the motion of charged particles in Kerr-Newman space-times by Xu \cite{Xu11}. Geodesics in even more complicated space-times like the one incorporated in the Pleba\'nski-Demia\'nski space-time are beyond the methods introduced by Hagihara. However, in 2008 a method has been found to analytically integrate the geodesic equation in Schwarzschild-de Sitter space-times using the theory of hyperelliptic functions \cite{HackmannLaemmerzahl08,HackmannLaemmerzahl08b} and to classify all types of possible orbits in terms of energy and angular momentum of the particles as well as the cosmological constant. This approach was then used to find solutions of the geodesic equations in general Pleba\'nski-Demia\'nski space-times with vanishing acceleration of the gravitating source \cite{Hackmannetal09} and to some special cases \cite{Hackmannetal2010,Kagramanovaetal2010}. 

Bound stable orbits far away from the black hole are not very different from Newtonian orbits but in the vicinity of the black holes they exhibit some peculiar features. These are caused by mismatches between the periodicities of the radial, polar, and azimuthal motion, which are all equal to $2\pi$ in the Keplerian case but may differ greatly near black holes. A systematic approach to handle these different periodicities for Kerr space-time was worked out by Schmidt \cite{Schmidt02} and continued by Drasco and Hughes \cite{DrascoHughes04} using a technique for decoupling the radial and polar motion suggested by Mino \cite{Mino03}. They showed that arbitrary functions of Kerr black hole orbits can be described in the frequency domain and how these expansions may be computed explicitly. Later, Fujita and Hikida \cite{FujitaHikida09} derived analytical expressions for the frequencies of radial, polar, and azimuthal motion in Kerr space-time and also for bound timelike orbits in Kerr space-time. 

In this paper, we will find analytic expressions of the fundamental frequencies defined by Schmidt \cite{Schmidt02}, Drasco and Hughes \cite{DrascoHughes04}, as well as Fujita and Hikida \cite{FujitaHikida09} for the general Pleba\'nski-Demia\'nski space-time with vanishing acceleration of the gravitating source. We will then investigate how these frequencies are influenced by the parameters of the black hole. To this end, we expand the analytic expression in terms of the parameters in a Taylor series up to first order. This clearly shows the influence of different parameters on the orbits of the black hole and leads to conclusions which kinds of orbits may belong to a given black hole.

The paper is organized as follows: First, we review the equations of motion in Pleba\'nski-Demia\'nski space-times decoupled by the method introduced by Mino. These equations are then used in the third section to define the fundamental frequencies and observables for Pleba\'nski-Demia\'nski space-time following closely the arguments of Fujita and Hikida. All necessary quantities in Schwarzschild space-time as well as their post-Newtonian expansions are computed in section four. These quantities will serve as a reference for the comparison with more complicated space-times later on. Section five is the most technical one, where the linear correction to the fundamental frequencies are computed. These corrections are given in terms of elementary function or in terms of complete Jacobian elliptic functions. In section six we use this results to compute the post-Schwarzschild and post-Newtonian corrections to the periastron shift, the Lense-Thirring effect, and the conicity and compare them to earlier results. A discussion and outlook closes the paper.

\section{Geodesic motion in Pleba\'nski-Demia\'nski space-time}\label{Sec:PDgeod}

The axially symmetric Pleba\'nski-Demia\'nski space-times are characterized by the seven parameters mass, rotation, acceleration, cosmological constant, NUT-parameter, electric and magnetic charge. It can be shown that the Hamilton-Jacobi equation for these space-times is separable and the geodesic equation integrable, if and only if the acceleration vanishes or null geodesics are considered. In this paper we will consider the case of geodesic motion of massive test-particles in a Pleba\'nski-Demia\'nski space-time with vanishing acceleration. We will also assume that the test-particles are neutral, i.e.~without electric or magnetic charge. 

The six-parameter Pleba\'nski-Demia\'nski space-times considered here are then given by the metric \cite{PlebanskiDemianski76,KubiznakKrtous07}\footnote{The metric (1) originates from \cite{KubiznakKrtous07} with $t \to t+2n\varphi$, $\theta \to \pi-\theta$} (we use units where $c=1=G$)
\begin{align}
ds^2/M^2 & = \frac{\Delta_r}{p^2} \left(dt - A d\varphi\right)^2 - \frac{p^2}{\Delta_r} dr^2  \nonumber\\
& \quad - \frac{\Delta_\theta}{p^2} \sin^2\theta (a dt - B d\varphi)^2 - \frac{p^2}{\Delta_\theta} d\theta^2 \, , \label{metric}
\end{align}
where $p^2 = r^2 + \left(n - a \cos\theta \right)^2$, $A = a \sin^2\theta  + 2 n \cos\theta$, $B = r^2 + a^2 + n^2$,
\begin{align}
\Dr & = (r^2 + a^2 - n^2)(1-\Lambda(r^2+3n^2))\nonumber\\
& \quad - 2r + Q^2_{\textrm e} + Q_{\textrm m}^2 - 4 \Lambda n^2r^2\,, \\
\Dt & = 1 + a^2 \Lambda \cos^2\theta - 4 \Lambda a n \cos\theta\,.
\end{align}
Here $a$, the angular momentum per mass of the gravitating source, the NUT parameter $n$, the electric charge $Q_{\rm e}$, and the magnetic charge $Q_{\rm m}$ as well as the coordinates $r$ and $t$ are normalized with respect to $M$, where $M$ is the mass of the gravitating object. The dimensionless parameter $\Lambda$ denotes the cosmological constant divided by three and normalized by multiplication with $M^2$.

The equations of motion for massive test-particles in these space-times are given by 
\begin{align}
\left( \frac{dr}{d\lambda} \right)^2 & = P(r)^2 - \Dr (r^2 + C) =: R(r) \label{reom} \\ 
\left( \frac{d\theta}{d\lambda} \right)^2 & = \Delta_\theta \left(C - (n - a \cos \theta)^2 \right) - \frac{O(\theta)^2}{\sin^2\theta}  =: \Theta(\theta) \label{thetaeom} \\ 
\frac{d\varphi}{d\lambda} & = \frac{a}{\Dr} P(r) + \frac{O(\theta)}{\Dt \sin^2\theta} =: \Phi(r,\theta) \label{phieom}\\
\frac{dt}{d\lambda} & = \frac{r^2+a^2+n^2}{\Dr} P(r) + \frac{a \sin^2\theta + 2n \cos\theta}{\Dt \sin^2\theta} O(\theta)\nonumber\\
& =: T(r,\theta) \label{teom}
\end{align}
where 
\begin{align}
P(r) & = (r^2 + a^2 + n^2) E - a L\,, \\
O(\theta) & = L - (a \sin^2\theta + 2n \cos\theta) E\,.
\end{align}

The constants of motion $E$, $L$, and $C$ have the meaning of energy, angular momentum in direction of the symmetry axes, and Carter constant, each per unit mass. The two constants $L$ and $C$ are additionally normalized by division by $M$ and $M^2$, respectively, such that they are dimensionless. The affine parameter $\lambda$ is the Mino time \cite{Mino03}, also normalized to $M$, $d\lambda = d\tau/(M p^2)$ with the eigentime $\tau$. Note that $R$ depends quadratically on $n$, $Q_{\textrm e}$, and $Q_{\textrm m}$ and that $\Theta$ does not depend on $Q_{\textrm e}$ and $Q_{\textrm m}$.

The equations \eqref{reom}-\eqref{teom} can be solved analytically \cite{Hackmannetal09,Kagramanova09,HackmannDiss}. However, in this paper we are interested in the periods of the motion, which can be used to define observables related to geodesic motion in these space-times.

\section{Observables for bound orbits}\label{Sec:Obs}

In Newtonian gravity bound geodesic motion is described by a fixed ellipse defining the orbital plane. This is no longer true in the framework of general relativity. In the spherically symmetric Schwarzschild space-time, the orbital plane remains fixed while the ellipse precesses resulting in a periastron shift. In the axially symmetric Kerr space-time additionally the orbital plane itself precesses what is known as the Lense-Thirring effect\footnote{With 'Lense-Thirring effect' we refer to the precession of the orbital plane only, not including the additional precession of the periapsis as derived in the original paper \cite{LenseThirring18}}. In the more exotic Taub-NUT space-time the geodesic motion of test particles does not lie in a plane at all but on a cone, see e.g.~\cite{LyndenNouri98}. 

The precession of the orbital ellipse and the orbital plane is induced by mismatches of the periods of the motion in the $r$ and $\theta$ coordinates compared to the average secular increase of the angle $\varphi$ about the symmetry axes. These effects were discussed in the framework of a Kerr space-time by Schmidt \cite{Schmidt02}, Drasco and Hughes \cite{DrascoHughes04}, and Fujita and Hikida \cite{FujitaHikida09}. In the following sections, we will use the procedure in \cite{FujitaHikida09} to analyze the first order corrections to the periastron shift and the Lense-Thirring effect due to the parameters $a$, $n$, and $\Lambda$. (As $Q_e$ and $Q_m$ appear only quadratically, there are no linear effects due to electric or magnetic charge on neutral test particles.) Below we review their line of argument for the convenience of the reader. In addition, we will characterize the deviation from an orbital plane, the conicity, due to the parameter $n$. 

For bound orbits, the radial and polar components $r$ and $\theta$ vary between a minimal and maximal value given by the turning points $\frac{dr}{d\lambda}=0$ and $\frac{d\theta}{d\lambda}=0$. The periods $\varLambda_r$ of $r$ and $\varLambda_\theta$ of $\theta$ with respect to the Mino time $\lambda$ are then defined by a revolution from maximum to minimum and back to the maximal value. This means that $\varLambda_r$ and $\varLambda_\theta$ are defined by the smallest non-zero real value with $r(\lambda+\varLambda_r)=r(\lambda)$ and $\theta(\lambda+\varLambda_\theta)=\theta(\lambda)$ giving
\begin{align}
\varLambda_r & = \oint_{\mathfrak{a}_r} \frac{dr}{\sqrt{R(r)}} = 2 \int_{r_{\textrm p}}^{r_{\textrm a}} \frac{dr}{\sqrt{R(r)}}\,,\\
\varLambda_\theta & = \oint_{\mathfrak{a}_\theta} \frac{d\theta}{\sqrt{\Theta(\theta)}} = 2 \int_{\theta_{\textrm min}}^{\theta_{\textrm max}} \frac{d\theta}{\sqrt{\Theta(\theta)}}\,,
\end{align}
where $r_{\textrm p}$ is the periapsis and $r_{\textrm a}$ the apoapsis. In general these are hyperelliptic integrals, which are a generalization of elliptic integrals. (The closed integration path $\mathfrak{a}_r$ ($\mathfrak{a}_\theta$) refers to the integration on the Riemann surface of the algebraic curve defined by $y^2=R(r)$ ($y^2=\Theta(\cos\theta)$). It runs around the branch cut connecting $r_{\textrm p}$ and $r_{\textrm a}$ ($\cos\theta_{\textrm min}$ and $\cos\theta_{\textrm max}$). On the Riemann surface the two branches of the square root are glued to one analytic function.) From these two periods conjugate fundamental frequencies can be defined by
\begin{equation}
\Upsilon_r := \sL \frac{2\pi}{\varLambda_r} \quad \,, \, \Upsilon_\theta := \sL \frac{2\pi}{\varLambda_\theta}\,.
\end{equation}
The sign of $L$ is included in this definition to indicate the direction in which the particle travels around the gravitating object relative to $a>0$. This means that $L>0$ corresponds to a prograde and $L<0$ to a retrograde orbit.

For vanishing $n$, Eq.~\eqref{thetaeom} is symmetric with respect to the equatorial plane $\theta = \frac{\pi}{2}$. However, in the more general space-time considered here this is not true, and the deviation from this symmetry can be measured by the difference between $\frac{1}{2}(\theta_{\textrm min}+\theta_{\textrm max})$ and $\frac{\pi}{2}$. In the case of a Taub-NUT space-time, where all parameters except mass and NUT parameter vanish, this implies that the particle moves on an orbital cone rather than an orbital plane \cite{MisnerTaub69}. Later we will see that this phenomenon also appears in the weak field limit. For these reasons, we will refer to this quantity as the conicity, $\Delta_{\textrm{conicity}}:=\pi-(\theta_{\textrm min}+\theta_{\textrm max})$. This means that $\Delta_{\textrm{conicity}}>0$ corresponds to a cone opened in northern direction and $\Delta_{\textrm{conicity}}<0$ to a cone opened in southern direction.

The nature of the equations \eqref{phieom} and \eqref{teom} is somewhat different from \eqref{reom} and \eqref{thetaeom} as they cannot be solved by periodic functions. They depend on both $r$ and $\theta$ but can be separated in an $r$-dependent and a $\theta$-dependent part
\begin{align}
\Phi(r,\theta) & =: \Phi_r(r) + \Phi_\theta(\theta)\,,\\
T(r,\theta) & =: T_r(r) + T_\theta(\theta)\,.
\end{align}
The solutions $\varphi(\lambda)$ and $t(\lambda)$ of equations \eqref{phieom} and \eqref{teom} can be written as an averaged part linear in $\lambda$ plus perturbations in $r$ and $\theta$
\begin{align}
\varphi(\lambda) & = \langle \Phi(r,\theta) \rangle_\lambda \lambda + \Phi^r_{\textrm osc}(r) + \Phi^\theta_{\textrm osc}(\theta) \\
t(\lambda) & = \langle T(r,\theta) \rangle_\lambda \lambda + T^r_{\textrm osc}(r) + T^\theta_{\textrm osc}(\theta)
\end{align}
where 
\begin{equation}
\langle \cdot \rangle_\lambda:=\lim_{(\lambda_2-\lambda_1) \to \infty} \frac{1}{2(\lambda_2-\lambda_1)} \oint_{\lambda_1}^{\lambda_2} \cdot \, d\lambda
\end{equation} 
is an infinite time average with respect to $\lambda$, and $X^r_{\textrm osc}(r)$ and $X^\theta_{\textrm osc}(\theta)$ (with $X=\Phi$ or $X=T$) represent oscillatory deviations from this average. They are defined by
\begin{align}
X^r_{\textrm osc}(r) & = \int X_r(r) d\lambda - \langle X_r(r) \rangle_\lambda \lambda \,, \\
X^\theta_{\textrm osc}(\theta) & = \int X_\theta(\theta) d\lambda - \langle X_\theta(\theta) \rangle_\lambda \lambda
\end{align}
and have periods $\varLambda_r$ and $\varLambda_\theta$. Therefore, the average secular increase of $\varphi$ and $t$ with respect to $\lambda$ is given by
\begin{align}
\Gamma & := \langle T(r,\theta) \rangle_\lambda = \langle T_r(r) \rangle_\lambda + \langle T_\theta(\theta) \rangle_\lambda\,,\\ \Upsilon_\varphi &:= \langle \Phi(r,\theta) \rangle_\lambda = \langle \Phi_r(r) \rangle_\lambda + \langle \Phi_\theta(\theta) \rangle_\lambda \,,
\end{align}
As $X_r$ and $X_\theta$ (again with $X=\Phi$ or $X=T$) are periodic functions with respect to $\lambda$ because $r$ and $\theta$ are periodic functions, their integrals $\oint_{\lambda_1}^{\lambda_2} X_r(r(\lambda)) d\lambda$ and $\oint_{\lambda_1}^{\lambda_2} X_\theta(\theta(\lambda)) d\lambda$ in the definition of the infinite time average can be reduced to an integral over one period $\varLambda_r$ or $\varLambda_\theta$, respectively. This yields
\begin{align}
\Upsilon_\varphi & = \frac{2}{\varLambda_r} \int_{r_{\textrm p}}^{r_{\textrm a}} \frac{\Phi_r(r) dr}{\sqrt{R(r)}} + \frac{2}{\varLambda_\theta} \int_{\theta_{\textrm min}}^{\theta_{\textrm max}} \frac{\Phi_\theta(\theta) d\theta}{\sqrt{\Theta(\theta)}}\,, \\
\Gamma & = \frac{2}{\varLambda_r} \int_{r_{\textrm p}}^{r_{\textrm a}} \frac{T_r(r) dr}{\sqrt{R(r)}} + \frac{2}{\varLambda_\theta} \int_{\theta_{\textrm min}}^{\theta_{\textrm max}} \frac{T_\theta(\theta) d\theta}{\sqrt{\Theta(\theta)}}\,.
\end{align}
The corresponding frequencies with respect to the coordinate time $t$ are then given by
\begin{align}
\Omega_r := \frac{\Upsilon_r}{\Gamma}\,, \quad \Omega_\theta := \frac{\Upsilon_\theta}{\Gamma}\,, \quad \Omega_\varphi := \frac{\Upsilon_\varphi}{\Gamma}\,.
\end{align}
These are frequencies as seen by an observer at infinity.

In the limit of weak gravitational fields the mismatch of the frequency of the $\varphi$- and $r$-motion, $\Omega_{\textrm P} = \Omega_\varphi - \Omega_r$ can be interpreted as the precession of the orbital ellipse, and the mismatch of the frequency of the $\varphi$- and $\theta$-motion, $\Omega_{\textrm{LT}} = \Omega_\varphi - \Omega_\theta$ as a precession of the orbital plane. In strong gravitational fields the orbits are in general of such irregularity that orbital planes or ellipses can no longer be identified.

\section{Reference orbit}\label{sec:reference}

In the next section we will analyze the influence of the parameters $a$, $\Lambda$, and $n$ on the observables of a bound reference orbit, which is neither circular nor polar. (Polar orbits may be considered as a special case but we leave this out here. For a discussion of these orbits in Kerr space-time see \cite{Kraniotis07}.) This will be done by post-Schwarzschild and post-Newton expansions, where we assume the constants of motions to be fixed. But first we will introduce and characterize the reference orbit.

If all space-time parameters except the mass vanish, the Pleba\'nski-Demia\'nski space-time is identically to the Schwarzschild space-time and the functions $R(r)$ and $\Theta(\theta)$ of Eqs.~\eqref{reom} and \eqref{thetaeom} reduce to 
\begin{align}
R_0(r) &:= (E^2-1)r^4+2r^3-Cr^2+2Cr\,, \\
\Theta_0(\theta) & := C-\frac{L^2}{\sin^2\theta}\,.
\end{align}
A necessary condition for the existence of a non-circular bound orbit in a Schwarzschild space-time is that $R_0(r)$ has four real zeros, $0=r_{01}<r_{02}<r_{03}<r_{04}<\infty$, where $R_0(r)>0$ for $r_{03}<r<r_{04}$, cf.~\cite{HackmannDiss}. Therefore, our bound reference orbit has turning points $r_{03}$ and $r_{04}$. Note that instead of using $E^2$ and $C$ all formulas can also be expressed in terms of the turning points: A comparison of coefficients in $(E^2-1)r^4+2r^3-Cr^2+2Cr=(E^2-1)r(r-r_{02})(r-r_{03})(r-r_{04})$ yields
\begin{align}
E^2 -1 & = \frac{-2}{r_{02}+r_{03}+r_{04}}\,, \label{defE} \\
C & = \frac{r_{02}r_{03}r_{04}}{r_{02}+r_{03}+r_{04}}\,, \label{defC} \\
r_{02} & = \frac{-2r_{03}r_{04}}{2r_{03}+2r_{04}-r_{03}r_{04}} \label{defr02} \,.
\end{align}
A non-polar orbit requires $C\geq L^2$ and lies in an orbital plane with inclination $\arcsin \frac{|L|}{\sqrt{C}}$ (or $\frac{\pi}{2}-\arcsin \frac{|L|}{\sqrt{C}}$ if measured from the equatorial plane). Therefore, the theta motion of the reference orbit is symmetric with respect to the equatorial plane and confined to $[\theta_{01},\theta_{02}]$ with $\theta_{01} = \arcsin \frac{|L|}{\sqrt{C}} \in [0,\frac{\pi}{2}]$ and $\theta_{02}=\pi-\arcsin \frac{|L|}{\sqrt{C}} \in [\frac{\pi}{2},\pi]$. In particular, the orbit lies in the equatorial plane if $C=L^2$. In terms of the inclination and the turning points the constant of motion $L$ is given by 
\begin{align}
L = \pm \sqrt{C} \sin \theta_{01} = \pm \sqrt{\frac{r_{02}r_{03}r_{04}}{r_{02}+r_{03}+r_{04}}} \sin \theta_{01}\,. \label{defL}
\end{align}
In the following we will calculate the expressions $\Upsilon_r$, $\Upsilon_\theta$, $\Upsilon_\phi$, and $\Gamma$ for the reference orbit.

\subsection{Frequency of $r$}\label{ref_rmotion}
Let us first calculate the $r$ period for our reference orbit. For an orbit bound between $r_{03}$ and $r_{04}$ we get 
\begin{align}\label{period_r}
\varLambda_{r,0} : = 2 \int_{r_{03}}^{r_{04}} \frac{dr}{\sqrt{R_0(r)}}\,.
\end{align}
This is a complete elliptic integral of first kind which can be easily transformed to the Legendre form giving
\begin{align}\label{r_Legendre}
\varLambda_{r,0} = \frac{4 K(k)}{\sqrt{(1-E^2)r_{03}(r_{04}-r_{02})}} \,,
\end{align}
with $k^2 = r_{02}(r_{04}-r_{03})/(r_{03}(r_{04}-r_{02}))$. For general information on the complete elliptic integral of first kind
\begin{align}
K(k) = \int_0^1 \frac{dt}{\sqrt{(1-t^2)(1-k^2 t^2)}}
\end{align}
see e.g.~\cite{AbramowitzStegun64}, for fast numerical computation see e.g.~\cite{Fukushima09}. In computer algebra systems like Mathematica or Maple the complete elliptic integrals are usually implemented and, therefore, $\varLambda_{r,0}$ can be computed easily. The conjugate fundamental frequency $\Upsilon_{r,0}$ is then given by
\begin{align}
\Upsilon_{r,0} = \frac{2 \pi}{4 K(k)} \sL \sqrt{(1-E^2)r_{03}(r_{04}-r_{02})}\,.
\end{align}

\subsection{Frequency of $\theta$}\label{ref_thetamotion}
The $\theta$ period of the reference orbit is given by
\begin{align}
\varLambda_{\theta,0} &:= 2 \int_{\theta_{01}}^{\theta_{02}} \frac{d\theta}{\sqrt{\Theta_0(\theta)}}\nonumber\\
& = 2 \int_{\xi_{02}}^{\xi_{03}} \frac{d\xi}{\sqrt{C(1-\xi^2)-L^2}}
\,,
\end{align}
where we substituted $\xi = \cos\theta$ and $\cos\theta_{01} = \sqrt{1-\frac{L^2}{C}} =: \xi_{03}$, $\cos\theta_{02} = \xi_{02} = -\xi_{03}$. This can be solved by 
\begin{align}
\varLambda_{\theta,0} & = \frac{2\pi}{\sqrt{C}} = \frac{2\pi \sqrt{r_{02}+r_{03}+r_{04}}}{\sqrt{r_{02}r_{03}r_{04}}}
\end{align}
from which we infer
\begin{align}
\Upsilon_{\theta,0} & = \sL \frac{2\pi}{\varLambda_{\theta,0}} = \sL \sqrt{C}
\,.
\end{align}
Note that for the case $\theta_{01}=\frac{\pi}{2}=\theta_{02}$ it is $\theta(\lambda) \equiv \frac{\pi}{2}$ and, therefore, $\varLambda_{\theta,0}$ is undefined. However, we can treat this as the limiting case $C\to L^2$, which gives the same results for $\varLambda_{\theta,0}$ and $\Upsilon_{\theta,0}$ as above.

\subsection{Frequency of $\varphi$}

For the reference orbit Eq.~\eqref{phieom} simplifies to
\begin{align}
\frac{d\varphi}{d\lambda} = \frac{L}{\sin^2\theta}
\end{align}
and $\Upsilon_\varphi$ can be calculated to
\begin{align}
\Upsilon_{\varphi,0} & = \frac{2}{\varLambda_{\theta,0}} \int_{\theta_{01}}^{\theta_{02}} \frac{L}{\sin^2\theta \sqrt{\Theta_0(\theta)}}\nonumber\\
&
= \frac{2\pi \sL}{\varLambda_{\theta,0}} = \sL \sqrt{C}\,,
\end{align}
where we used the substitution $\xi=\cos\theta$ as in sec.~\ref{ref_thetamotion}. Although $\varLambda_{\theta,0}$ is not defined for $\theta(\lambda) \equiv \frac{\pi}{2}$ we get in this case $\varphi(\lambda) = L\lambda$ and, therefore, the same result $\Upsilon_{\varphi,0} = L = \sL \sqrt{C}$.

\subsection{Frequency of $t$}\label{ref_tmotion}

The expression for $\Gamma_0$ is the most complicated in this section as it involves an elliptic integral of third kind. For the reference orbit Eq.~\eqref{teom} simplifies to
\begin{align}
\frac{dt}{d\lambda} & = \frac{r^3 E}{r-2}\,,
\end{align}
which leads to
\begin{align}
\Gamma_0 & = \frac{2}{\varLambda_{r,0}} \int_{r_{02}}^{r_{03}} \frac{r^3 E dr}{(r-2) \sqrt{R_0(r)}}\\
& = \frac{E}{2 K(k)} \bigg( -r_{03} r_{04} K(k) + r_{03}(r_{04}-r_{02}) E(k) \nonumber\\
& \quad + \frac{2r_{04}(3-2E^2)}{1-E^2} \Pi(n_1,k) + \frac{8 r_{04}}{r_{04}-2} \Pi(n_2,k) \bigg)\,, \nonumber
\end{align}
where $E(k)$ and $\Pi(n,k)$ are the complete elliptic integrals of second and third kind,
\begin{align}
E(k) & = \int_0^1 \frac{(1-k^2t^2) dt}{\sqrt{(1-t^2)(1-k^2t^2)}}\,, \\
\Pi(n,k) & = \int_0^1 \frac{dt}{(1-nt^2)\sqrt{(1-t^2)(1-k^2t^2)}}\,,
\end{align}
and the parameters $n_1$, $n_2$ are given by
\begin{align}\label{def_n}
n_1 = \frac{r_{03}-r_{04}}{r_{03}} <0 \,, \qquad n_2 = \frac{2(r_{04}-r_{03})}{r_{03}(r_{04}-2)} <0\,.
\end{align}

\subsection{Observables} \label{Ref_Obs}

Let us now collect the results so far in this section. 
As expected for $a=0$, the frequencies $\Upsilon_{\theta,0}$ and $\Upsilon_{\varphi,0}$ coincide and the Lense-Thirring effect vanishes
\begin{align}
\Omega_{\textrm{LT},0} & = \frac{\Upsilon_{\varphi,0}-\Upsilon_{\theta,0}}{\Gamma_0} = 0\,.
\end{align}
Likewise, for $n=0$ the particle moves on an orbital plane, 
\begin{align}
\Delta_{\textrm{conicity},0} = \pi - (\theta_{01}+\theta_{02})\comment{\pi - \left(\arcsin \frac{L}{\sqrt{C}} + \pi - \arcsin \frac{L}{\sqrt{C}} \right)} = 0\,.
\end{align}
The periastron shift $\Omega_{{\textrm P},0} = (\Upsilon_{\varphi,0}-\Upsilon_{r,0})\Gamma_0^{-1}$ is given in terms of elliptic integrals
\begin{widetext}
\begin{align} \label{OmegaP0}
\Omega_{{\textrm P},0} & = \frac{\sL}{E} \frac{2 \sqrt{C} K(k) - \pi \sqrt{r_{03}(r_{04}-r_{02})(1-E^2)}}{\left(-r_{03}r_{04}K(k) + r_{03}(r_{04}-r_{02}) E(k) + \frac{2r_{04}(3-2E^2)}{1-E^2} \Pi(n_1,k) + \frac{8r_{04}}{r_{04}-2} \Pi(n_2,k) \right)}\,.
\end{align}
\end{widetext}
The usual result for the periastron shift in terms of radians or degrees can be found in the following way: using the averaged $\lambda(\varphi) = \varphi \Upsilon_{\varphi}^{-1}$ we get the period of the $r$-motion in terms of $\varphi$, $r_\varphi:= r \circ \lambda: \varphi \mapsto r(\varphi \Upsilon_{\varphi}^{-1})$, by observing that $r_\varphi(\varphi + \Upsilon_{\varphi} \varLambda_{r}) = r( \varphi \Upsilon_{\varphi}^{-1} + \varLambda_{r}) = r_\varphi(\varphi)$. From this period we get for the difference between the angle of the periapsis and $2\pi$ after one revolution
\begin{align}\label{DeltaP0}
\Delta_{{\textrm P},0} & = \Upsilon_{\varphi,0} \varLambda_{r,0} - 2\pi\sL\\
& = \sL \left(  \frac{4\sqrt{C}K(k)}{\sqrt{(1-E^2)r_{03}(r_{04}-r_{02})}} -2\pi \right). \nonumber
\end{align}
If this shift should be expressed in terms of radians per time, it can be referred to the time needed for a revolution from some fixed $\varphi_0$ to $\varphi_0$ again, i.e.~the sidereal period, or for a revolution from some fixed $r_0$ to $r_0$ again, i.e.~the anomalistic period. The first choice corresponds to the usual notion of the periastron shift in arcseconds per century whereas the second corresponds to the definition of $\Omega_{\textrm P}$: the time elapsed for a revolution from, say, periapsis to periapsis is given as the period of $r_t:= r \circ \lambda: t \mapsto r(t \Gamma^{-1})$, which is $\Gamma \varLambda_r$. If $\Delta_{\textrm P} = \Upsilon_{\varphi} \varLambda_r - 2\pi \sL$ is divided by this period we obtain $\Omega_{\textrm P} = \Delta_{\textrm P}/(\Gamma \varLambda_r)$.

Note that \eqref{OmegaP0} and \eqref{DeltaP0} are exact and, therefore, more complicated than the post-Newtonian formula given e.g.~in \cite{Will06}. If we consider the weak field approximation by assuming that the periapsis $r_{03}$ and the apoapsis $r_{04}$ become large, we recover the formula (51) of \cite{Will06} with vanishing solar quadrupole momentum and PPN parameters corresponding to general relativity,
\begin{align}
\Delta_{{\textrm P},0} \approx \frac{6\pi \sL}{d(1-\epsilon^2)} M \,,
\end{align}
where $d=\frac{M}{2}(r_{\textrm a}+r_{\textrm p})$ is the semimajor axis and $\epsilon=\frac{M}{2d}(r_{\textrm a}-r_{\textrm p})$ the eccentricity with the apoapsis $r_{\textrm a}=r_{04}$ and the periapsis $r_{\textrm p}=r_{03}$. Here we used \eqref{defE}-\eqref{defr02} to perform the series expansion. Note that the angular momentum of the test particle is usually chosen to be positive and, therefore, its sign does not appear in Eq.~(51) of \cite{Will06}. But we include it here as we choose the sign of $L$ relative to the angular momentum of the gravitating source, which will be nonzero later on. The post-Newtonian expression for $\Omega_{\textrm P}$ is given by
\begin{align}
\Omega_{{\textrm P},0} \approx \frac{3 \sL}{d^\frac{5}{2} (1-\epsilon^2)} M^\frac{5}{2}\,,
\end{align}
which includes the perturbation of the time needed for one revolution from periapsis to periapsis.

Let us test our results for the orbital motion of Mercury. From \cite{Horizons} we take for peri- and aphelion the values
\begin{align}
r_{\textrm min} & = (307500.7 \pm 3.0) \times 10^{-6} \, \textrm{AU}\,, \\
r_{\textrm max} & = (466696.6 \pm 2.6) \times 10^{-6} \, \textrm{AU}\,,
\end{align}
and use $M = 1476.625\,28 \, \textrm m$ for the mass of the Sun to determine $r_{03}=\frac{r_{\textrm min}}{M}$ and $r_{04}=\frac{r_{\textrm max}}{M}$. With Eqs.~\eqref{defE}-\eqref{defr02} and the results of this section we obtain
\begin{align}
\Omega_{\textrm{P},0} & = 3.252 308 \times 10^{-19} \pm 5.5 \times 10^{-24} 
\end{align}
In order to express this dimensionless quantity in terms of arcseconds per century we have to multiply it by $\frac{c}{M}$ where $c=299 792 458 \, \textrm{m} \,\textrm{s}^{-1}$ is the speed of light. Then we get
\begin{align}\label{OmegaP0_num}
\Omega_{{\textrm P},0} \frac{c}{M} & = 42.980\, 48 \pm (0.73 \times 10^{-3}) \, \, \textrm{arcsec}/\textrm{cy}\,, 
\end{align}
in consistency with observations, cf.~\cite{Pireauxetal03}. However, it is not the usual result in the sense that it describes the perihelion shift per revolution from periapsis to periapsis rather than from $0$ to $2\pi$.

The usual result for the perihelion shift in terms of arcseconds per century can by found using $\Delta_{\textrm P}$,
\begin{align}
\Delta_{{\textrm P},0} & = 5.018\, 648\, 5 \times 10^{-7} \pm 3.7 \times 10^{-12}\,.
\end{align}
In \cite{Horizons} a revolution of Mercury is given as $87.969257$ days. We get with $t_\varphi:=t\circ \lambda: \varphi \mapsto \Gamma_0 \Upsilon_{\varphi,0}^{-1} \varphi$ for a revolution of $2\pi$
\begin{align}\label{Mercuryyear}
{\textrm y}_{\textrm M} := \frac{2\pi \Gamma_0}{\Upsilon_{\varphi,0}} \, \frac{M}{c} = 87.969\, 25 \pm (8.5 \times 10^{-4})  \, \textrm{days}\,,
\end{align}
which agrees to the given accuracy with observation. This yields for the perihelion shift
\begin{align}
\frac{\Delta_{{\textrm P},0}}{{\textrm y}_{\textrm M}} & = 42.980\, 48 \pm (0.73 \times 10^{-3}) \, \, \textrm{arcsec}/\textrm{cy}\,,
\end{align}
in accord with observations. We see here that this value coincides with \eqref{OmegaP0_num} within the given accuracy and, therefore, the two different definitions can not be distinguished.

\section{First order corrections}\label{sec:firstorder}

In the following we will calculate the linear post-Schwarzschild corrections for all quantities used to define the observables $\Omega_{\textrm P}$ and $\Omega_{\textrm{LT}}$ as well as the conicity $\Delta_{\textrm{conicity}}$ due to the parameters $a$, $n$, and $\Lambda$. As the parameters $Q_{\textrm e}$ and $Q_{\textrm m}$ appear only quadratically in Eqs.~\eqref{reom} to \eqref{teom}, we will not study them here. However, this would be a totally analogous procedure. We assume the constants of motion to be fixed but let the zeros $r_i$ of $R$ and $\theta_i$ of $\Theta$ vary. By this procedure we will reduce the hyperelliptic integrals appearing in the definitions of $\varLambda_r$, $\varLambda_\theta$, $\Upsilon_\varphi$, and $\Gamma$ for the general Pleba\'nski-Demia\'nski space-time to elliptic integrals and elementary expressions.

The observables $\Omega_{\textrm P}$ and $\Omega_{\textrm{LT}}$ are defined through the frequencies $\Upsilon_r$, $\Upsilon_\theta$, $\Upsilon_\varphi$, and $\Gamma$. Whereas $\Upsilon_r$ and $\Upsilon_\theta$ depend only on the variable indicated in the index, $\Upsilon_\varphi$ and $\Gamma$ can be separated in an $r$- and a $\theta$-dependent part, $\Upsilon_\varphi = \Upsilon_{\varphi r} + \Upsilon_{\varphi \theta}$ and $\Gamma=\Gamma_r+\Gamma_\theta$, where
\begin{align}
\begin{aligned} \label{def:Iphi}
\Upsilon_{\varphi r} & := \frac{2}{\varLambda_r} \int_{r_p}^{r_a} \frac{\Phi_r(r) dr}{\sqrt{R(r)}} =: \frac{I_{\varphi r}}{\varLambda_r} \,,\\
\Upsilon_{\varphi \theta} & := \frac{2}{\varLambda_\theta} \int_{\theta_{\textrm min}}^{\theta_{\textrm max}} \frac{\Phi_\theta(\theta) d\theta}{\sqrt{\Theta(\theta)}} =: \frac{I_{\varphi \theta}}{\varLambda_\theta}\,,
\end{aligned}\\
\begin{aligned}\label{def:Gamma}
\Gamma_r & := \frac{2}{\varLambda_r} \int_{r_p}^{r_a} \frac{T_r(r) dr}{\sqrt{R(r)}} =: \frac{I_{t r}}{\varLambda_r} \,,\\
\Gamma_\theta & := \frac{2}{\varLambda_\theta} \int_{\theta_{\textrm min}}^{\theta_{\textrm max}} \frac{T_\theta(\theta) d\theta}{\sqrt{\Theta(\theta)}} =: \frac{I_{t \theta}}{\varLambda_\theta}\,. 
\end{aligned}
\end{align}
In terms of these expressions the periastron shift and the Lense-Thirring effect can be restated as
\begin{align}
\Omega_{\textrm P} & = \frac{\Upsilon_\varphi-\Upsilon_r}{\Gamma} = \frac{I_{\varphi r}\varLambda_\theta+I_{\varphi \theta}\varLambda_r- \sL 2\pi\varLambda_\theta}{I_{t r}\varLambda_\theta+I_{t \theta}\varLambda_r}\,, \label{Periastron}\\
\Omega_{\textrm{LT}} & = \frac{\Upsilon_\varphi-\Upsilon_\theta}{\Gamma} = \frac{I_{\varphi r}\varLambda_\theta+I_{\varphi \theta}\varLambda_r- \sL 2\pi\varLambda_r}{I_{t r}\varLambda_\theta+I_{t \theta}\varLambda_r}\,. \label{LenseThirring}
\end{align}

\subsection{Standard form of hyperelliptic integrals} \label{sec:standard}

All integrals appearing here have the form
\begin{align}
\int_{x_1}^{x_2} \frac{f(x) dx}{\sqrt{P_6(x)}}\,, \label{int_start}
\end{align}
where $P_6$ is a polynomial of degree 6 in $x=r$ or $x=\xi=\cos\theta$, $f(x)$ is a rational function, and $x_1$, $x_2$ are zeros of $P_6$. This type of integral is called hyperelliptic integral. The functions $P_6$ and $f$ as well as the zeros $x_1$, $x_2$ depend on the parameters $a$, $n$, and $\Lambda$. For a Taylor expansion of such an integral it is of advantage to first reduce it to a standard form similar to the elliptic integrals which appeared in the calculations for the reference orbit. However, to our knowledge such a standard form does not exist in the literature. A straightforward generalization of the Legendre standard form of elliptic integrals of first kind can be obtained by an additional term $(1-k_2^2t^2)$ under the square root but this yields again an elliptic integral as a substitution $s=t^2$ shows. A better choice is to generalize the Riemann form of elliptic integrals,
\begin{align}
\int_0^1 \frac{dt}{\sqrt{t(1-t)(1-k^2t)}}\,,
\end{align}
to the hyperelliptic form
\begin{align}
\int_0^1 \frac{(At+B) dt}{\sqrt{t(1-t)(1-k_1^2t)(1-k_2^2t)(1-k_3^2t)}}\,, \label{standard}
\end{align}
which was also used in \cite{Kraniotis07} for the calculation of the periastron shift of equatorial orbits in Kerr-de Sitter space-times. As pointed out in \cite{Kraniotis07}, the form \eqref{standard} can also be expressed in terms of Lauricella's hypergeometric $F_D$ function, see also appendix \ref{app:FD}. 

The transformation of the form \eqref{int_start} to the form \eqref{standard} depends on the range of integration $[x_1,x_2]$. In the case of $P_6(x)=R(r)$, we have 4 real zeros $r_1<r_2<r_3<r_4$ and another two zeros $r_0,r_5$ which may be complex and which tend to infinity for vanishing $\Lambda$. Here $x_1=r_3$ and $x_2=r_4$ result in a transformation $r=(At+B)^{-1} +r_1$ which yields
\begin{align}
&\int_{r_3}^{r_4} \frac{dr}{\sqrt{R(r)}}\nonumber\\
& = \frac{1}{\sqrt{D}} \int_0^1 \frac{(At+B) dt}{\sqrt{t(1-t)(1-k_1^2t)(1-k_2^2t)(1-k_3^2t)}} \label{FDform} \\
& = \frac{A}{\sqrt{D}} \frac{\pi}{2} F_D\left(\frac{3}{2},\vec{\beta}_1,2,\vec{m}\right) + \frac{B}{\sqrt{D}} \pi F_D\left(\frac{1}{2},\vec{\beta}_1,1,\vec{m}\right)\,, \nonumber
\end{align}
where $B=1/(r_4-r_1)=:B_r$,
\begin{align}
A & = \frac{r_4-r_3}{(r_4-r_1)(r_3-r_1)} =: A_r\,,\\
D & = \Lambda \frac{(r_3-r_1)(r_4-r_2)}{(r_4-r_1)^2} (r_4-r_0)(r_5-r_4) =: D_r\,, \\
k_1^2 & = \frac{(r_4-r_3)(r_2-r_1)}{(r_3-r_1)(r_4-r_2)} =: k_{1r}^2\,,\\
k_2^2 & = - \frac{(r_4-r_3)(r_5-r_1)}{(r_3-r_1)(r_5-r_4)} =: k_{2r}^2\,,\\
k_3^2 & = - \frac{(r_4-r_3)(r_1-r_0)}{(r_3-r_1)(r_4-r_0)} =: k_{3r}^2\,,
\end{align}
and $\vec{\beta}_1=\left(\frac{1}{2},\frac{1}{2},\frac{1}{2}\right)$, $\vec{m}=(k_{1r}^2,k_{2r}^2,k_{3r}^2)=:\vec{m}_{0r}$. For the case of $P_6(x)=\Theta_\xi(\xi)$ (see Eq.~\eqref{def_Thetaxi}) there are two real zeros $\xi_2<\xi_3$ and four other maybe complex zeros $\xi_0$, $\xi_1$, $\xi_4$, $\xi_5$, where $\xi_1$, $\xi_4$ tend to infinity for vanishing $a$ and $\xi_0$, $\xi_5$ for vanishing $a$ or $\Lambda$. With $x_1=\xi_2$ and $x_2=\xi_3$ we get the same form \eqref{FDform} but with $B=1/(\xi_3-\xi_1)=:B_\xi$,
\begin{align}
A & = \frac{(\xi_3-\xi_2)}{(\xi_2-\xi_1)(\xi_3-\xi_1)} =: A_\xi\,,\\
D & = a^4\Lambda \frac{(\xi_2-\xi_1)(\xi_4-\xi_3)}{(\xi_3-\xi_1)^2} (\xi_3-\xi_0)(\xi_3-\xi_5) =: D_\xi\,, \\
k_1^2 & = -\frac{(\xi_3-\xi_2)(\xi_4-\xi_1)}{(\xi_2-\xi_1)(\xi_4-\xi_3)} =: k_{1\xi}^2\,,\\
k_2^2 & = - \frac{(\xi_3-\xi_2)(\xi_5-\xi_1)}{(\xi_2-\xi_1)(\xi_5-\xi_3)} =: k_{2\xi}^2\,,\\
k_3^2 & = - \frac{(\xi_3-\xi_2)(\xi_1-\xi_0)}{(\xi_2-\xi_1)(\xi_3-\xi_0)} =: k_{3\xi}^2\,,
\end{align}
and $\vec{m}=(k_{1\xi}^2,k_{2\xi}^2,k_{3\xi}^2)=:\vec{m}_{0\xi}$.

\subsection{Linear correction to period of $r$}\label{rmotion}

In this section we will calculate the linear corrections to the reference orbit due to the parameters $a$, $n$, and $\Lambda$. The exact formula for the $r$ period is given by
\begin{align}
\varLambda_r & = 2 \int_{r_3}^{r_4} \frac{dr}{\sqrt{R(r)}} \nonumber\\
& = \frac{2}{\sqrt{D_r}} \int_0^1 \frac{(A_rt+B_r) dt}{\sqrt{t(1-t)(1-k_{1r}^2t)(1-k_{2r}^2t)(1-k_{3r}^2t)}} \nonumber\\
& = \frac{2A_r}{\sqrt{D_r}} \frac{\pi}{2} F_D\left(\frac{3}{2},\vec{\beta}_1,2,\vec{m}_{0r}\right) \nonumber \\
& \qquad + \frac{2B_r}{\sqrt{D_r}} \pi F_D\left(\frac{1}{2},\vec{\beta}_1,1,\vec{m}_{0r}\right)\,. \label{rperiod_exact}
\end{align}
All quantities indexed with an $r$ are given as some combination of the zeros of $R(r)$ and, therefore, depend on the parameters $a$, $n$, and $\Lambda$. This means that for a Taylor expansion of $\varLambda_r$ we need to explicitly know how the zeros $r_i$ of $R(r)$ depend on the parameters. To see this, we compare the coefficients of the equation
\begin{align}
R(r) & = \Lambda \left( r-\frac{\tilde{r}_0}{\sqrt{\Lambda}}\right) (r-r_1)(r-r_2)\nonumber\\
& \quad \times (r-r_3)(r-r_4)\left(r-\frac{\tilde{r}_5}{\sqrt{\Lambda}}\right)\,, \label{Rcoeffs}
\end{align}
where $\tilde{r}_0=r_0\sqrt{\Lambda}$, $\tilde{r}_5 = r_5 \sqrt{\Lambda}$ do not have a singularity at $\Lambda=0$. As the coefficient of $r^5$ vanishes in $R(r)$ we immediately see that 
\begin{align}
-(r_2+r_3+r_4+r_1) \Lambda - (\tilde{r}_0+\tilde{r}_5) \sqrt{\Lambda} = 0
\end{align}
and, thus, $\tilde{r}_0$ and $\tilde{r}_5$ have to be expanded in terms of $\sqrt{\Lambda}$. Therefore, we introduce $l=\sqrt{\Lambda}$ as expansion parameter instead of $\Lambda$. Expanding the right hand side of \eqref{Rcoeffs} in terms of $a$, $n$, and $l$ gives a system of equations which can be solved for the derivatives of all $r_i$ with respect to the parameters.

The Taylor expansion of $\varLambda_r$ in $\vec{p}=(a,n,l)$ near $\vec{p}=\vec{0}$ then reads
\begin{widetext}
\begin{align}
\varLambda_r & \approx \varLambda_{r}(\vec{p}=\vec{0}) + \sum_{i=1}^3 \, \varLambda_{r,i} (\vec{p}=\vec{0}) \,\, p_i \\
& = \varLambda_{r,0} + 2 \sum_{i=1}^3 \bigg\{ \frac{-D_{r,i}}{2 D_r^\frac{3}{2}} \left[ A_r \frac{\pi}{2} F_D\left(\frac{3}{2},\vec{\beta}_1,2,\vec{m}_{0r}\right) + B_r \pi F_D\left(\frac{1}{2},\vec{\beta}_1,1,\vec{m}_{0r}\right) \right] \nonumber\\
& \qquad + \frac{1}{\sqrt{D_r}} \left[ A_{r,i} \frac{\pi}{2} F_D\left(\frac{3}{2},\vec{\beta}_1,2,\vec{m}_{0r}\right) + B_{r,i} \pi F_D\left(\frac{1}{2},\vec{\beta}_1,1,\vec{m}_{0r}\right) \right] + \frac{1}{\sqrt{D_r}} \sum_{j=1}^3 \frac{(k_{jr}^2)_{,i}}{2k_{jr}^2} \left[  A_r \frac{\pi}{2} \left( F_D\left(\frac{3}{2},\vec{\beta}_1^j,2,\vec{m}_{0r}\right) \right. \right. \nonumber\\
& \qquad \qquad \left. \left. - F_D\left(\frac{3}{2},\vec{\beta}_1,2,\vec{m}_{0r}\right) \right) + B_r \pi \left( F_D\left(\frac{1}{2},\vec{\beta}_1^j,1,\vec{m}_{0r}\right) - F_D\left(\frac{1}{2},\vec{\beta}_1,1,\vec{m}_{0r}\right) \right) \right] \bigg\}\bigg|_{\vec{p}=\vec{0}} \, p_i \,, \label{rtaylor}
\end{align}
where we used $\frac{\partial}{\partial x_i} F_D(\alpha,\vec{\beta},\gamma,\vec{x}) = \frac{\beta_i}{x_i} (F_D(\alpha,\vec{\beta}^i,\gamma,\vec{x}) - F_D(\alpha,\vec{\beta},\gamma,\vec{x}))$ which can be found in \cite{Miller72}. The symbol $\vec{\beta}^j$ means that the $j$th component of $\vec{\beta}$ is increased by one (e.g.~$\vec{\beta}_1^{2}=(\frac{1}{2},\frac{3}{2},\frac{1}{2})$), and by $X_{,i}$ we denote the derivative of $X$ with respect to $p_i$. Note that
\begin{align}
k_{2r}^2 & =\frac{r_{03}-r_{04}}{r_{03}}=k_{3r}^2\,, \qquad \text{for } \vec{p}=\vec{0}\,,
\end{align}
and, therefore, the Lauricella function $F_D$ reduces to an elliptic integral in this case. Here $r_{0i}$ are again the zeros of $R_0$, i.e.~the turning points of the reference orbit.

For the linear correction with respect to $a$ we obtain
\begin{align}
\varLambda_{r,a}(\vec{0}) & = -4EL \frac{\zeta_1 (1-E^2)^2(r_{03}-r_{02}) K(k) + 2r_{03} (4-3C+3CE^2) E(k)}{(1-E^2)^3\sqrt{r_{03}(r_{04}-r_{02})(1-E^2)}(r_{04}-r_{02})(r_{03}-r_{02})^2(r_{04}-r_{03})^2}\,,
\end{align}
where $k^2 = \frac{r_{02}(r_{04}-r_{03})}{r_{03}(r_{04}-r_{02})}$ as for the reference orbit and $\zeta_1$ is a non-symmetric function of $r_{02}$, $r_{03}$, and $r_{04}$,
\begin{align}\label{defzeta1}
\zeta_1 & = r_{02}r_{03}+r_{02}r_{04}-r_{03}^2-r_{04}^2\,.
\end{align}
The linear corrections with respect to $n$ vanishes,
\begin{align}
\varLambda_{r,n}(\vec{0}) = 0
\end{align}
and also the linear correction due to $l=\sqrt{\Lambda}$, as expected. To determine the linear correction due to $\Lambda$ we use
\begin{align}
\frac{df}{d\Lambda}\left(a,n,\Lambda\right)\big|_{\Lambda=0} = \left[ \frac{df}{dl}(a,n,l) \, \frac{1}{2l} \right]_{l=0}\,.
\end{align}
In this way the linear correction due to $\Lambda$ can be calculated to
\begin{align}
\varLambda_{r,\Lambda}(\vec{0}) & = \frac{1}{\sqrt{r_{03}(r_{04}-r_{02})(1-E^2)}(1-E^2)} \bigg[ \frac{r_{03}r_{04}(\gamma_1-2C\gamma_2)}{(r_{02}-r_{04})(r_{03}-r_{04})^2(r_{02}-r_{03})} K(k) \nonumber \\
& \qquad - \frac{4 r_{03} C (-16C+76CE^2-57CE^4-C^2+C^2E^4-48)}{(r_{02}-r_{04})(r_{02}-r_{03})^2(r_{03}-r_{04})^2(1-E^2)^5} E(k) + \frac{6r_{04}}{(1-E^2)^2} \Pi(n_1,k) \bigg]
\end{align}
\end{widetext}
where $\gamma_1$ and $\gamma_2$ are non-symmetric functions of $r_{02}$, $r_{03}$, and $r_{04}$,
\begin{align}\label{defgamma}
\begin{split}
\gamma_1 & = 2r_{03}^2r_{04}^2 -3r_{03}^3r_{04} -3r_{04}^3r_{03} +3r_{03}^3r_{02}\\
& \quad  +3r_{04}^3r_{02} -r_{03}^2r_{02}^2 +2r_{03}r_{04}r_{02}^2 -r_{04}^2r_{03}r_{02}\\
& \quad -r_{04}^2r_{02}^2 -r_{03}^2r_{04}r_{02}\,, \\
\gamma_2 & = r_{04}r_{02}+r_{03}r_{02}-2r_{04}r_{03}\,.
\end{split}
\end{align}

\subsection{Linear correction to period of $\theta$}\label{thetamotion}
The linear correction to the $\theta$ period can be determined analogously to the foregoing subsection. The exact formula for the period of the $\theta$ motion is
\begin{align}
\varLambda_\theta &= \frac{2A_\xi}{\sqrt{D_\xi}} \frac{\pi}{2} F_D\left(\frac{3}{2},\vec{\beta}_1,2,\vec{m}_{0\xi}\right) \nonumber\\
& \quad + \frac{2B_\xi}{\sqrt{D_\xi}} \pi F_D\left(\frac{1}{2},\vec{\beta}_1,1,\vec{m}_{0\xi} \right)\,. \label{thetaperiod_exact}
\end{align}
Again, we have to determine the dependence of the zeros $\xi_i$, $i=0,\ldots,5$ on the parameters $a$, $n$ and $\Lambda$. This time we use the ansatz
\begin{align}
\Theta_\xi(\xi) &= a^4 \Lambda (\xi-\xi_2)(\xi-\xi_3) \left(\xi-\frac{\tilde{\xi}_1}{a}\right) \left(\xi-\frac{\tilde{\xi}_4}{a}\right) \nonumber\\
& \quad \times \left(\xi-\frac{\tilde{\xi}_0}{a\sqrt{\Lambda}}\right) \left(\xi-\frac{\tilde{\xi}_5}{a\sqrt{\Lambda}}\right) \label{Thetacoeffs}
\end{align}
where $\Theta_\xi(\xi)$ is the right hand side of \eqref{thetaeom} with the substitution $\xi=\cos\theta$,
\begin{align}
\left(\frac{d\xi}{d\lambda}\right)^2 & = \Theta_\xi(\xi) \nonumber\\
& = (1+a^2\Lambda \xi^2-4\Lambda an\xi)(C-(n-a\xi)^2) (1-\xi^2)\nonumber\\
& \quad -(L-(a(1-\xi^2)+2n\xi)E)^2\,, \label{def_Thetaxi}
\end{align}
and $\tilde{\xi}_1=a\xi_1$, $\tilde{\xi}_4=a\xi_4$, $\tilde{\xi}_0=a\sqrt{\Lambda}\xi_0$, $\tilde{\xi}_5=a\sqrt{\Lambda}\xi_5$ behave regular in the limit $a=0$, $\Lambda=0$. By comparing the coefficient of $\xi^5$ it can be seen that $\xi_0$ and $\xi_5$ expand in terms of $l=\sqrt{\Lambda}$ like $r_0$ and $r_5$ in the foregoing section. Solving the system of equations given by \eqref{Thetacoeffs} we obtain $\tilde{\xi}_0 \approx -i \sqrt{1-E^2}$, $\tilde{\xi}_5 \approx i\sqrt{1-E^2}$ and, therefore, $\tilde{\xi}_0$ and $\tilde{\xi}_5$ are complex conjugate.

For the Taylor expansion of $\varLambda_\theta$ we obtain the same formula \eqref{rtaylor} with $r$ replaced by $\xi$. However, in this case we get
\begin{align}
k_{1\xi}^2=k_{2\xi}^2=k_{3\xi}^2 = 0 \quad \text{for } \vec{p}=0\,,
\end{align}
which means that the Lauricella function $F_D$ reduces to an elementary function in the limit $\vec{p}=0$. For the linear correction with respect to $a$ we get
\begin{align}
\varLambda_{\theta,a}(\vec{0}) = - \frac{2\pi LE}{C^{\frac{3}{2}}}\,.
\end{align}
The corrections with respect to $n$ and $l$ vanish, $\varLambda_{\theta,n}(\vec{0}) = 0 = \varLambda_{\theta,l}(\vec{0})$, as well as the correction for $\Lambda$ due to $\varLambda_{\theta,l}(\vec{0})=\mathcal{O}(l^2)$.

\subsection{Linear correction to frequency of $\varphi$ motion}\label{phimotion}

In this section we will calculate the Taylor expansions of the integrals $I_{\varphi r}$ and $I_{\varphi \theta}$ defined in \eqref{def:Iphi}, 
\begin{align*}
I_{\varphi r}(\vec{p}) & = 2 \int_{r_p}^{r_a} \frac{\Phi_r(r) dr}{\sqrt{R(r)}}\,, \quad I_{\varphi \theta}(\vec{p}) = 2 \int_{\theta_{\textrm min}}^{\theta_{\textrm max}} \frac{\Phi_\theta(\theta) d\theta}{\sqrt{\Theta(\theta)}}\,.
\end{align*}
The conversion of these two integrals to the standard form is more involved than in the case of $\varLambda_r$ and $\varLambda_\theta$ because an additional function $\Phi_r$ or $\Phi_\theta$ appears in the integrand here. Let us first consider $\Phi_r(r) = a \frac{ P(r)}{\Dr}$. The poles of $\Phi_r$ are given by the horizons $\Dr = 0$, which we denote by $h_i$, $1\leq i \leq 4$. Let $h_2 \approx 0$ and $h_3 \approx 2$ correspond to the Schwarzschild case, whereas $h_1 \approx -l^{-1} -1$ and $h_4 \approx l^{-1} -1$ tend to infinity for vanishing $\Lambda$. Then $\Phi_r$ can be rewritten as 
\begin{widetext}
\begin{align}
\Phi_r(r) = \frac{a P(r)}{l^2\left(r-\frac{\tilde{h}_1}{l}\right)(r-h_2)(r-h_3)\left(r-\frac{\tilde{h}_4}{l}\right)}
\end{align}
where $\tilde{h}_1 = lh_1$, $\tilde{h}_4 = lh_4$. Now $\Phi_r$ can be decomposed in partial fractions and the resulting integrals transformed to the standard form by the substitution outlined in section \ref{sec:standard}. This yields an integral of the form
\begin{align}
I_{\varphi r}(\vec{p}) & = 2 \sum_{i=0}^4 \frac{c_{i \varphi r}}{\sqrt{D_r}} \int_0^1 \frac{(A_rt+B_r) dt}{(1-N_{i \varphi r}t)\sqrt{t(1-t)(1-k_{1r}^2t)(1-k_{2r}^2t)(1-k_{3r}^2t)}}  \\
& = 2 \sum_{i=0}^4 \frac{c_{i \varphi r}}{\sqrt{D_r}} \left[ A_r \frac{\pi}{2} F_D\left(\frac{3}{2},\vec{\beta}_2,2,\vec{m}_{i \varphi r}\right) + B_r \pi F_D\left(\frac{1}{2},\vec{\beta}_2,1,\vec{m}_{i \varphi r}\right) \right]\,, \label{Iphir}
\end{align}
where $c_{i \varphi r}$ and $N_{i \varphi r}$ are some constants with $N_{0 \varphi r}=0$, $\vec{\beta}_2 = \left(\frac{1}{2},\frac{1}{2},\frac{1}{2},1\right)$, and $\vec{m}_{i \varphi r} = (k_{1r}^2,k_{2r}^2,k_{3r}^2,N_{i \varphi r})$. The Taylor expansion of this can be done analogously to section \ref{rmotion} and gives
\begin{align}
I_{\varphi r} (\vec{p}) \approx \frac{4 E r_{04}\, \Pi(n_2,k)}{(r_{04}-2)\sqrt{r_{03}(r_{04}-r_{02})(1-E^2)}} \, a
\end{align}
as the other corrections with respect to $n$, $l$, and $\Lambda$ vanish. Here $r_{0i}$, $k$, and $n_2$ correspond to the reference orbit, see section \ref{sec:reference} for a definition.

Now let us consider $I_{\varphi \theta}$. With the substitution $\xi=\cos\theta$ the integral can be transformed to
\begin{align}
I_{\varphi \theta} (\vec{p})
& = 2 \int_{\xi_2}^{\xi_3} \frac{(L-(a(1-\xi^2)+2n\xi)E) \, d\xi}{(\xi-1)(\xi+1)\left(\xi-\frac{1}{l}(2nl+f(n,l))\right)\left(\xi-\frac{1}{l}(2nl-f(n,l))\right) \sqrt{\Theta_\xi(\xi)}}
\end{align}
where $f(n,l)=\sqrt{4l^2n^2-1}$. Then the standard form can be obtained with the substitution described in section \ref{sec:standard} giving the integral \eqref{Iphir} with $r$ replaced by $\xi$. Here again $c_{i \varphi \xi}$ and $N_{i \varphi \xi}$ are constants with $N_{0 \varphi \xi}=0$
. The Taylor expansion is given by
\begin{align}
I_{\varphi \theta} (\vec{p}) \approx 2\pi \sL - \frac{2\pi E}{\sqrt{C}} \, a\,,
\end{align}
with vanishing corrections due to $n$, $l$, and $\Lambda$.

\subsection{Linear correction to frequency of $t$ motion}

The two integrals to be considered for determining the correction to the $t$ motion are
\begin{align*}
I_{t r} (\vec{p}) & = 2 \int_{r_p}^{r_a} \frac{T_r(r) dr}{\sqrt{R(r)}}\,,\quad I_{t \theta}(\vec{p}) = 2 \int_{\theta_{\textrm min}}^{\theta_{\textrm max}} \frac{T_\theta(\theta) d\theta}{\sqrt{\Theta(\theta)}}\,.
\end{align*}
The procedure will be analogous to the foregoing subsection. The poles of $T_r$ and $T_\theta$ are the same as of $\Phi_r$ and $\Phi_\theta$ respectively and, therefore, the integrals can be expressed similar to \eqref{Iphir}. We get
\begin{align}
I_{t r}(\vec{p}) & = 2 \sum_{i=0}^4 c_{i t r} \bigg[ A_r \frac{\pi}{2} F_D\left(\frac{3}{2},\vec{\beta}_2,2,\vec{m}_{i t r}\right) + B_r \pi F_D\left(\frac{1}{2},\vec{\beta}_2,1,\vec{m}_{i t r}\right) \bigg]\,, \label{Itr} \\
& \approx I_{t r}(\vec{0}) + I_{t r,a}(\vec{0}) a + I_{t r,n}(\vec{0}) n + I_{t r,\Lambda}(\vec{0}) \Lambda \nonumber
\end{align}
where $c_{itr}$ and $N_{itr}$ are constants with $N_{0tr}=0$, $\vec{\beta}_2=(\frac{1}{2},\frac{1}{2},\frac{1}{2},1)$, $\vec{m}_{itr}=(k_{1r}^2,k_{2r}^2,k_{3r}^2,N_{itr})$, and $I_{tr}(\vec{0})=I_{t r,0}$ as for the reference orbit. The linear correction due to $a$ is given by
\begin{align}
I_{t r,a}(\vec{0}) & = \frac{4Lr_{03}^2}{(r_{03}(r_{04}-r_{02})(1-E^2))^\frac{3}{2} (r_{04}-r_{03})^2 (r_{03}-r_{02})} \bigg[ \frac{r_{04} \eta_1 E^2 K(k)}{(r_{04}-2)(r_{03}-2)} + \frac{2 C(8+2C-3CE^2)}{(r_{03}-r_{02})(1-E^2)^2} E(k) \bigg]\,,
\end{align}
where $\eta_1$ is a non-symmetric function of $r_{02}$, $r_{03}$, and $r_{04}$,
\begin{align}\label{defeta1}
\eta_1 & = 2r_{04}^2r_{02}-r_{03}r_{04}^2r_{02}-2r_{04}^2r_{03} +2r_{03}^2r_{04}^2-2r_{03}^2r_{04}-r_{03}^2r_{04}r_{02}+2r_{03}^2r_{02}\,.
\end{align}
The correction due to $n$ vanishes, $I_{t r,n}(\vec{0})=0$, and the correction due to $\Lambda$ is given by 
\begin{align}
& I_{t r,\Lambda}(\vec{0}) = \frac{E}{\sqrt{r_{03}(r_{04}-r_{02})(1-E^2)}} \nonumber\\
& \times \Bigg\{ -r_{03}r_{04} K(k) \bigg[ \left( \frac{(\alpha_2 +12C\alpha_3)}{24(1-E^2)(r_{04}-r_{03})^2(r_{03}-r_{02})(r_{04}-r_{02})} + \frac{\alpha_1}{12} \right) (r_{04}-2)^{-1}(r_{03}-2)^{-1} + \frac{2}{(1-E^2)^3} \bigg] \nonumber  \\
& \quad + \frac{r_{03}(r_{04}-r_{02})}{(1-E^2)} E(k) \bigg[ \frac{(\alpha_4+12C\alpha_5)}{24S} - \frac{8-75E^2+149E^4-116E^6+32E^8}{E^2(1-E^2)^2} - \frac{4C}{3}  \bigg] \nonumber \\
&\quad + \frac{r_{04}}{1-E^2} \Pi(n_1,k) \bigg[ \frac{238-754E^2+952E^4-560E^6+2^7E^8}{(1-E^2)^3} + \frac{2C(2E^2-5)}{1-E^2} + \frac{\alpha_6+4C\alpha_7}{8(r_{04}-r_{03})^2(r_{03}-r_{02})(r_{04}-r_{02})} \bigg] \nonumber \\
&\quad + \frac{r_{02}+r_{03}-3r_{04}}{(1-E^2)^2} \bigg[ r_{03}r_{04} K(k) - r_{03}(r_{04}-r_{02}) E(k) - \frac{2r_{04}(3-2E^2)}{1-E^2} \Pi(n_1,k) \bigg] + \frac{256r_{04}}{r_{04}-2} \Pi(n_2,k) \Bigg\}\,, 
\end{align}
\end{widetext}
where $S=(r_{04}-r_{03})^2(r_{03}-r_{02})^2(r_{04}-r_{02})^2(r_{04}-2)(r_{03}-2)(r_{02}-2)$ and $\alpha_i$, $i=1,\ldots,7$, are non-symmetric functions of $r_{02}$, $r_{03}$, and $r_{04}$ given in appendix \ref{app:OmegaPLambda}.

The Taylor expansion for the $\theta$ dependent part is much easier. The zero order term and all linear correction except $I_{t \theta,a}$ vanish, $I_{t \theta}(\vec{0}) =I_{t \theta,n}(\vec{0})=I_{t \theta,\Lambda}(\vec{0})=0$. The remaining term is given by
\begin{align}
I_{t \theta,a}(\vec{0}) & = \frac{2\pi L}{\sqrt{C}}\,.
\end{align}

\section{Observables}
In section \ref{Sec:Obs} we defined the general expressions for the fundamental frequencies $\Omega_r$, $\Omega_\theta$, and $\Omega_\varphi$ in Pleba\'nski-Demia\'nski space-time. The analytic expressions of these quantities are given in terms of complete hyperelliptic integrals on a Riemann surface of genus 2 (see Eqs.~\eqref{rperiod_exact}, \eqref{thetaperiod_exact}, \eqref{Iphir}, and \eqref{Itr}), whose numerical evaluation is quite cumbersome. By a post-Schwarzschild expansion of the fundamental frequencies we will reduce the hyperelliptic integrals to elementary expressions and elliptic integrals, which can be handled easily with computer algebra systems like Mathematica or Maple. In addition, we will perform the post-Newtonian expression for comparison with other results and use the parameter values of Mercury to get an idea of the order of magnitude of the corrections.

\subsection{Periastron shift}
In section \ref{sec:reference} we calculated the expression for $\Omega_{\textrm P}$ for the Schwarzschild case, see \eqref{OmegaP0},
and noted that as long as a particle is considered to move in an orbital plane usually the difference angle $\Delta_{\textrm P}$, see \eqref{DeltaP0}, is used instead of $\Omega_{\textrm P}$.
Let us now consider the linear corrections of $\Omega_{\textrm P}$ and $\Delta_{\textrm P}$ due to $a$, $n$, and $\Lambda$. As all linear corrections due to $n$ so far vanished, the linear correction due to this parameter also vanishes for $\Omega_{\textrm P}$ and $\Delta_{\textrm P}$. The linear correction due to $a$ for the frequency $\Omega_{\textrm P}$ is given by
\begin{widetext}
\begin{align}
\Omega_{{\textrm P},a}(\vec{0}) & = \frac{I_{\varphi r,a}(\vec{0}) + \sL \sqrt{C} \varLambda_{r,a}(\vec{0}) - E\varLambda_{r,0}\left(1-\frac{|L|}{\sqrt{C}}\right)}{I_{t r,0}} + \sL \frac{( L\varLambda_{r,0} + I_{t r,a}(\vec{0})) (2\pi-\sqrt{C}\varLambda_{r,0})}{I_{t r,0}^2} \nonumber \\
& = \frac{2}{Z} \bigg[ \frac{r_{04}}{r_{04}-2} \Pi(n_2,k) + \sqrt{C}|L| \frac{\zeta_1(r_{02}-r_{03})(E^2-1)^2 K(k) - 2r_{03}(4-3C(E^2-1))E(k)}{(r_{02}-r_{03})^2(r_{02}-r_{04})(r_{03}-r_{04})^2(E^2-1)^3} - K(k) \left(1-\frac{|L|}{\sqrt{C}}\right) \bigg] \nonumber \\
& \quad + \frac{|L|}{E^2 Z^2} \bigg[ K(k) + \left(\frac{r_{03}r_{04}\eta_1E^2 K(k)}{(r_{04}-2)(r_{03}-2)} + r_{03} \frac{2C(8+2C-3CE^2)E(k)}{(r_{03}-r_{02})(1-E^2)^2}\right) \nonumber \\
& \qquad \quad \times \left((1-E^2)(r_{04}-r_{02})(r_{04}-r_{03})^2(r_{03}-r_{02})\right)^{-1} \bigg] \bigg[ 2\pi \sqrt{r_{03}(r_{02}-r_{04})(E^2-1)} - 4 \sqrt{C} K(k) \bigg]
\end{align}
where $Z=-r_{03}r_{04}K(k)+r_{03}(r_{04}-r_{02})E(k)+2r_{04}\frac{3-2E^2}{1-E^2} \Pi(n_1,k) + \frac{8r_{04}}{r_{04}-2} \Pi(n_2,k)$ and $\zeta_1$, $\eta_1$ are non-symmetric functions of $r_{02}$, $r_{03}$, and $r_{04}$ defined in \eqref{defeta1} and \eqref{defzeta1}. Here $r_{0i}$, $k$, and $n_i$ correspond to the reference orbit and are defined in section \ref{sec:reference}. Note that $\Omega_{{\textrm P},a}$ does not depend on the sign of $L$, i.e.~on whether the particle travels on a prograde or retrograde orbit.

The correction of $\Delta_{\textrm P}$ due to the angular momentum of the gravitating source reads
\begin{align}
\Delta_{{\textrm P},a}(\vec{0}) & = \frac{4E\sqrt{C}}{\sqrt{(1-E^2)r_{03}(r_{04}-r_{02})}} \Bigg\{ \bigg[ \frac{|L|\zeta_1}{(r_{04}-r_{03})^2(r_{03}-r_{02})(r_{04}-r_{02})(E^2-1)}  - \frac{1}{\sqrt{C}} +\frac{|L|}{C} \bigg] K(k) \nonumber \\
& \quad  - \frac{2|L|r_{03}(4-3C-3CE^2)}{(1-E^2)^3(r_{04}-r_{03})^2(r_{03}-r_{02})^2(r_{04}-r_{02})} E(k)  + \frac{r_{04}}{r_{04}-2} \Pi(n_2,k) \Bigg\}\,,
\end{align}
\end{widetext}
which also does not depend on the sign of $L$. These formulas are exact in $M$ and, therefore, quite complicated. If we consider the post-Newtonian approximation of these terms by using \eqref{defE}-\eqref{defL} and assuming that $r_{03}$ and $r_{04}$, the turning points of the reference orbit in Schwarzschild space-time, become large we obtain
\begin{align}
\Omega_{{\textrm P},a}(\vec{0}) a & = 2 \, \frac{1-6\sin\theta_{01}}{d^3(1-\epsilon^2)^\frac{3}{2}} \, J M^2 + \mathcal{O}(J M^3)\,, \\
\Delta_{{\textrm P},a}(\vec{0}) a & = 4\pi \, \frac{1-6\sin\theta_{01}}{d^\frac{3}{2}(1-\epsilon^2)^\frac{3}{2}} \, J M^{\frac{1}{2}} + \mathcal{O}(J M^\frac{3}{2}) \,,
\end{align}
where $d$ is again the semimajor axis, $\epsilon$ the eccentricity, and $J = aM$ is the non-normalized specific angular momentum of the gravitating source. These terms are similar to the expression derived by Lense and Thirring \cite{LenseThirring18} for the precession of the longitude of pericenter, to which it differs only in the second term containing the inclination $\theta_{01}$. However, the fundamental frequency $\Omega_{\textrm P}$ is not identical to the precession of longitude of the pericenter: As $\Omega_{r}$ does not contribute to the first order correction of $\Omega_{\textrm P}$ the value of $\Omega_{{\textrm P}, a}$ is calculated by the correction due to $\Omega_\varphi = \frac{\Upsilon_\varphi}{\Gamma}$ only, which is the averaged $\frac{d\varphi}{dt}$. The longitude of pericenter on the other hand contains the argument of the pericenter which is defined not on the reference plane but the orbital plane.

Note that the expressions for $\Omega_{\textrm P}$ and $\Delta_{\textrm P}$ depend on the inclination of the orbital plane. For an inclination of $\theta_{01} < \arcsin \frac{1}{6} \approx 0.053 \pi$ the periastron shift is perturbed in the direction of the rotation of the gravitating source whereas for $\theta_{01} > \arcsin \frac{1}{6}$ the perturbation is in the opposite direction. (Here an inclination of $\theta_{01}=\frac{\pi}{2}$ corresponds to the equatorial plane and $\theta_{01} \leq \frac{\pi}{2}$ by definition, see sec.~\ref{sec:reference}.) In particular, if the unperturbed test particle moves on the equatorial plane, the periastron shift is perturbed against the direction of rotation. This seems to be counter-intuitive as a particle radially approaching the gravitational source is dragged along the direction of the rotation. However, this does not mean that the shape of the orbit is effected in the same way: the rotation of the gravitating source acts as an repulsive force which also causes the peri- and apoapsis to increase, see \eqref{reom}.

Let us now consider the linear correction of $\Omega_{\textrm P}$ and $\Delta_{\textrm P}$ due to $\Lambda$. For $\Omega_{\textrm P}$ we get a very complicated expression given in appendix \ref{app:OmegaPLambda}. In terms of the expression derived in the previous section we get
\begin{align}
\Omega_{{\textrm P},\Lambda} & = \sL \sqrt{C} \left[ \frac{\varLambda_{r,\Lambda}}{I_{t r,0}} + \left( \frac{2\pi}{\sqrt{C}} - \varLambda_{r,0} \right) \frac{I_{t r,\Lambda}}{I_{t r,0}^2} \right].
\end{align}
The correction of $\Delta_{\textrm P}$ due to $\Lambda$ is much simpler,
\begin{widetext}
\begin{align}\label{DeltaP_Lambdacorr}
\Delta_{{\textrm P},\Lambda} & = \frac{\sL \sqrt{C}}{\sqrt{(1-E^2)r_{03}(r_{04}-r_{02})}(1-E^2)} \Bigg\{ \frac{2C(r_{02}r_{03}+r_{02}r_{04}-2r_{03}r_{04})-\gamma_1}{(r_{04}-r_{03})^2(r_{04}-r_{02})(r_{03}-r_{02})} r_{03}r_{04} K(k) \nonumber \\
& \quad + \frac{48+16C-76CE^2+57CE^4+C^2-C^2E^4}{16+8C-36CE^2+27CE^4+C^2-C^2E^2} r_{03}(r_{04}-r_{02}) E(k) + \frac{6r_{04}}{1-E^2} \Pi(n_1,k) \Bigg\}\,.
\end{align}
\end{widetext}
These complicated expressions can be simplified by considering the post-Newtonian approximation which reads
\begin{align}
\Omega_{{\textrm P},\Lambda} \Lambda & = \frac{1}{2} \sL d^\frac{3}{2} \sqrt{1-\epsilon^2} \, \hat{\Lambda} M^\frac{1}{2} + \mathcal{O}(\hat{\Lambda} M^\frac{3}{2})\,, \label{OmegaPLambda_PN}
\end{align}
\begin{align}
\Delta_{{\textrm P},\Lambda} \Lambda & = \pi \sL d^3 \sqrt{1-\epsilon^2} \, \frac{\hat{\Lambda}}{M}\nonumber\\
& \quad  + 2\pi \sL d^2 \frac{2-\epsilon^2}{\sqrt{1-\epsilon^2}} \, \hat{\Lambda} + \mathcal{O}(\hat{\Lambda}M)\,, \label{DeltaP_Lambda}
\end{align}
where $\hat{\Lambda} = \frac{3\Lambda}{M^2}$ is the usual non-normalized cosmological constant. The first term of Eq.~\eqref{DeltaP_Lambda} coincides with the result in \cite{Kerretal03}. Interestingly, although we assumed both $\hat{\Lambda}$ and $M$ to be small we assumed nothing about their ratio. This raises the question whether the first term of \eqref{DeltaP_Lambda} is indeed the largest of the expansion. But if we divide the second term by the first this is proportional to $\frac{M}{d}$, which is assumed to be small. This proves that the terms of the above expansions in fact decrease in magnitude for increasing powers of $M$.

The exact formula for the perihelion shift assuming a non-vanishing cosmological constant was derived in \cite{KraniotisWhitehouse03,HackmannLaemmerzahl08b}. We will use the formalism of \cite{HackmannLaemmerzahl08b} (see Eq.~$(70)$) to compute the perihelion shift of Mercury in Schwarzschild-de Sitter space-times and compare it with our approximate formula. For this, we use $M = 1476.625\,28 \, \textrm m$ and, by using \eqref{defE}-\eqref{defL}, we determine averaged values for the energy and angular momentum from the apo- and periapsis data given by \cite{Horizons} and written down in \ref{Ref_Obs}. With the exact formulas given in \cite{HackmannLaemmerzahl08b} the difference between the perihelion shift for $\Lambda=0$ and with a cosmological constant of $\hat{\Lambda} = 3\Lambda/M^2 = 3 \times 10^{-52} \textrm m^{-2}$ is given by
\begin{multline}\label{DeltaP_exactLambda}
\Delta_{\textrm{P},(\Lambda=10^{-52})}/\textrm{y}_{\textrm M} - \Delta_{\textrm{P},(\Lambda=0)}/\textrm{y}_{\textrm M}\\
= 1.038\, 833\, 075\, 425\, 928 \times 10^{-14} \, \, \frac{\textrm{arcsec}}{\textrm{cy}}
\end{multline}
where ${\textrm y}_{\textrm M}$ is the Mercury year as calculated in \eqref{Mercuryyear}. If we insert the same values in our formula \eqref{DeltaP_Lambdacorr} we get
\begin{multline}
\Delta_{\textrm{P}}/\textrm{y}_{\textrm M} - \Delta_{\textrm{P},0}/\textrm{y}_{\textrm M} \\
= 1.038\, 832\, 108\, 177\, 831 \times 10^{-14}  \, \, \frac{\textrm{arcsec}}{\textrm{cy}} + \mathcal{O}(\Lambda^2)
\end{multline}
which agrees very well with \eqref{DeltaP_exactLambda}.

\subsection{Lense-Thirring effect}
The Lense-Thirring effect, which can be identified with a precession of the orbital plane in the weak field limit, is up to first order not influenced by any other parameter than the rotation $a$. This is because all linear corrections to $\Upsilon_\theta$ and $\Upsilon_\varphi$ due to $n$ and $\Lambda$ vanish. The linear correction due to $a$ is given by
\begin{align}
\Omega_{{\textrm{LT}},a} & = \frac{1}{Z} \left[\frac{2r_{04}}{(r_{04}-2)} \Pi(n_2,k)  -2K(k)\right]\,,
\end{align}
where $Z=-r_{03}r_{04}K(k)+r_{03}(r_{04}-r_{02})E(k)+2r_{04}\frac{3-2E^2}{1-E^2} \Pi(n_1,k) + \frac{8r_{04}}{r_{04}-2} \Pi(n_2,k)$ and $r_{0i}$, $k$, and $n_i$ correspond to the reference orbit, see section \ref{sec:reference}.

Let us consider now the weak field limit by assuming that $r_{03}$ and $r_{04}$ become large. Using \eqref{defE}-\eqref{defL} we get
\begin{align}
\Omega_{{\textrm{LT}},a} a & = \frac{2}{d^3(1-\epsilon^2)^\frac{3}{2}} \, JM^2 + \mathcal{O}(JM^3)\,,
\end{align}
where $J=aM$ is the non-normalized angular momentum of the gravitating source per unit mass, $d$ is the semimajor axis, and $\epsilon$ the eccentricity. This formula is identical to the precession rate of the longitude of the ascending node as given by Lense and Thirring, cf.~\cite{LenseThirring18}.

\subsection{Conicity}
The effect of $n$ can be observed by a deviation from the symmetry of the geodesic motion with respect to equatorial plane. In sec.~\ref{Sec:Obs} we defined $\Delta_{\textrm{conicity}}=\pi-(\theta_{\textrm min}+\theta_{\textrm max})$ as a measure for this deviation, which is given by
\begin{align}\label{1stconicity}
\Delta_{\textrm{conicity}} & \approx \sL \frac{4E}{\sqrt{C}} \, n 
\end{align}
up to first order. Linear corrections due to $a$ and $\Lambda$ vanish. This deviation from the symmetry to the equatorial plane also implies that for $n \neq 0$ a test particle can not move on an orbital plane and, in particular, not in the equatorial plane (if $E \neq 0$). Instead it moves on a cone with opening angle $\pi-\frac{4E}{\sqrt{C}}n + \mathcal{O}(n^2)$. As $\theta_{01}$ and $\theta_{02}$ are perturbed by the same value $\alpha=-\sL \frac{2E}{\sqrt{C}} n$ (up to first order) the symmetry axis of this cone coincides with one of the two normals of the orbital plane in the Schwarzschild case, i.e.~$\theta_{n}=\theta_{02}-\frac{\pi}{2} \in (0,\frac{\pi}{2})$ if $\alpha<0$ or $\theta_{n}=\theta_{01}+\frac{\pi}{2} \in (\frac{\pi}{2},\pi)$ if $\alpha>0$.

In the post-Newtonian limit the conicity becomes
\begin{align}
\Delta_{{\textrm{conicity}}} \approx \frac{4\sL}{\sqrt{d(1-\epsilon^2)}} \, \frac{\hat{n}}{M^\frac{1}{2}} - \frac{8\sL}{d^\frac{3}{2}(1-\epsilon^2)^\frac{3}{2}} \, \hat{n} M^\frac{1}{2} \,,
\end{align}
where $\hat{n} = nM$ is the non-normalized NUT parameter. Here $d$ and $\epsilon$ are the semimajor axis and the eccentricity, as before. Note that as we assumed $n=\frac{\hat{n}}{M}$ to be small in the first place, the first term in the above equation will indeed tend to zero for $M \to 0$ (but is still the largest of the expansion).

The conicity can be used to determine an upper bound for the value of $n$. As the error tolerance of the inclination of an orbiting object places a bound for $\Delta_{\textrm{conicity}}$, this can be used in one of the above equations to get an estimate of $n$. From \cite{Horizons} we obtain for the difference of maximal and minimal inclination of Mercury over a time span of 10 years $|\Delta_{\textrm{conicity}}| \leq 4.2001\, \textrm{arcsec}$ or $|\Delta_{\textrm{conicity}}| \leq 2.04 \times 10^{-5} \,{\rm rad}$. Inserted in \eqref{1stconicity} this yields $|n|\leq 0.032$.

\section{Conclusions and outlook}\label{Sec:Outlook}
In this paper we derived analytic expressions for the observables periastron shift and Lense-Thirring effect assuming neutral test particles in the general axially symmetric Pleba\'nski-Demia\'nski electrovac space-time with vanishing acceleration of the gravitating source. We determined the direct dependence of these observables on the parameters of the space-time by expanding them in a Taylor series up to first order. In addition, we defined the conicity of an orbit and analyzed its dependence on the space-time parameters by also expanding it in a Taylor series up to first order. Within the linear approximation we used the osculating orbital elements of Mercury to derive an upper bound for the dimensionless NUT parameter $n$. 

From the six parameters of a Pleba\'nski-Demia\'nski space-time with vanishing acceleration, only the Mass $M$, the rotation $a$, the NUT charge $n$, and the cosmological constant $\Lambda$ have any linear effects on neutral test particles. For Schwarzschild space-time ($0=a=n=\Lambda=Q_{\textrm e}=Q_{\textrm m}$), the only effect is the periastron shift, whereas the Lense-Thirring effect vanishes and the mean value of the polar coordinate coincides with the equator. In the following we summarize the effects due to the other parameters compared to Schwarzschild space-time:
\begin{itemize}
\item Taub-NUT space-time ($0=a=\Lambda=Q_{\textrm e}=Q_{\textrm m}$): The mean value of the polar motion deviates from the equator and the motion takes place on a cone rather than a plane.
\item Kerr space-time ($0=n=\Lambda=Q_{\textrm e}=Q_{\textrm m}$): The periastron shift is changed and the Lense-Thirring effect is nonzero. The latter can be interpreted as a precession of the orbital plane in the weak field approximation. Both are independent from the direction of rotation of the particle.
\item Kerr-Taub-NUT space-time ($0=\Lambda=Q_{\textrm e}=Q_{\textrm m}$): In addition to Kerr the mean value of the polar motion deviates from the equator. The combination with a non-vanishing Lense-Thirring effect causes a precession of the orbital cone.
\item Kerr-Taub-NUT-de Sitter space-time ($0=Q_{\textrm e}=Q_{\textrm m}$): In addition to Kerr-Taub-NUT the periastron shift is changed by a non-vanishing $\Lambda$. However, due to the smallness of $\Lambda$ the effect is tiny.
\end{itemize}
The effects of non-vanishing space-time parameters may also be analyzed compared to any other space-time covered by the general Pleba\'nski-Demia\'nski metric using the same methods as presented in this paper. In particular the Kerr space-time may be used as reference. In this case $\Upsilon_{\theta,0}$ and $\Upsilon_{\varphi,0}$ are also given in terms of elliptic integrals instead of elementary expressions.

The mathematical framework presented in this paper is based on the theory of hyperelliptic and elliptic functions. We defined a standard form for hyperelliptic integrals based on the Riemann form of elliptic integrals and expressed it in terms of the hypergeometric $F_D$ function introduced by Lauricella (see also \cite{Kraniotis07}). The methods presented in this paper may also be used to analyze the linear (and higher order) effects of the six parameters of Pleba\'nski-Demia\'nski space-time on charged particles with a complete analogous procedure. Another interesting question is whether additional effects arise for a higher order approximation, for example the quadratic effect of $n$ on $r$-dependent expressions or the coupling of $a$ to the other parameters. Similarly to the analysis presented here observables for unbound orbits like light deflection or the deflection of massive particles may be considered. This will be postponed to a later publication. 

\vspace*{3ex}
\begin{acknowledgments}
We are grateful to V.~Enolskii, V.~Kagramanova, J.~Kunz, P.~Richter, V.~Perlick, and N.~Wex for helpful discussions. E.H.~thanks the German Research Foundation DFG for financial support. C.L.~acknowledges support of the center of excellence QUEST.
\end{acknowledgments}

\appendix

\section{Lauricella's $F_D$ function}\label{app:FD}
The four functions $F_A$, $F_B$, $F_C$, and $F_D$ of Lauricella are hypergeometric functions of multiple variables generalizing the hypergeometric functions of Gauss and Appell. They were introduced in 1893 \cite{Lauricella1893} and given as a hypergeometric series
\begin{align}
F_D(\alpha,\vec{\beta},\gamma,\vec{x}) & = \sum_{\vec{\iota}=0}^{\infty} \frac{(\alpha)_{|\vec{\iota}|}\,(\vec{\beta})_{\vec{\iota}}}{(\gamma)_{|\vec{\iota}|}\, \vec{\iota}!} \,\, \vec{x}^{\vec{\iota}}\,,
\end{align}
where $\vec{\iota}$ is a multi-index, $|x_{\iota_n}|<1$ for all $n$, and $(\cdot)$ the Pochhammer symbol. Here $|\vec{\iota}|=\sum_n \iota_n$, $\vec{\iota}!=\prod_n \iota_n!$, and $(\vec{\beta})_{\vec{\iota}} = \prod_n (\beta_n)_{\iota_n}$. The function $F_D$ can be extended to other values of $\vec{x}$ by analytic continuation.

In this paper the $F_D$ function is used because it can be represented in an integral form
\begin{align}
& F_D(\alpha,\vec{\beta},\gamma,\vec{x}) = \frac{\Gamma(\gamma)}{\Gamma(\alpha)\Gamma(\gamma-\alpha)} \nonumber\\
& \quad \times \int_0^1 t^{\alpha-1} (1-t)^{\gamma-\alpha-1} \prod_n (1-x_nt)^{-\beta_n} dt 
\end{align}
for ${\textrm Re}(\gamma)>{\textrm Re}(\alpha)>0$, which is exactly the form of all hyperelliptic integrals appearing in this paper.

\begin{widetext}
\section{Details of $\Omega_{{\textrm P},\Lambda}$}\label{app:OmegaPLambda}
The calculation of the linear correction of the periastron shift due to the cosmological constant yields a very cumbersome expression given by
\begin{align}
\Omega_{{\textrm P},\Lambda} & = \frac{2E\sL \sqrt{C}}{(E^2-1) Z} \,  \bigg[\frac{\gamma_1-2C(r_{02}r_{03}+r_{02}r_{04}-2r_{03}r_{04})}{(r_{02}-r_{04})(r_{02}-r_{03})(r_{03}-r_{04})^2} r_{03}r_{04}K(k) \nonumber \\
& \qquad \quad + \frac{-16C+76CE^2-57CE^4-C^2+C^2E^4-48}{16+8C-36CE^2+27CE^4+C^2-C^2E^2}r_{03}(r_{04}-r_{02})E(k) + \frac{6r_{04}}{E^2-1} \Pi(n_1,k)\bigg] \nonumber \\
& \quad - \frac{\sL}{2EZ^2} \bigg[ 2\sqrt{C} K(k) - \pi\sqrt{r_{03}(r_{02}-r_{04})(E^2-1)} \bigg] \nonumber \\
& \quad \times \Bigg\{ \frac{r_{02}+r_{03}-3r_{04}}{(E^2-1)^2} \bigg[ r_{03}r_{04}K(k) - r_{03}(r_{04}-r_{02})E(k) -2r_{04}\frac{3-2E^2}{1-E^2} \Pi(n_1,k) \bigg] \nonumber \\
& \quad \quad + \bigg[ \frac{2}{(E^2-1)^3} - \frac{\alpha_1}{12(r_{04}-2)(r_{03}-2)} \nonumber \\
& \qquad \qquad + \frac{\alpha_2+12C\alpha_3}{24(r_{04}-2)(r_{03}-2)(r_{02}-r_{04})(r_{02}-r_{03})(r_{03}-r_{04})^2(E^2-1)} \bigg] r_{03}r_{04} K(k) \nonumber \\
& \quad \quad + \bigg[ \frac{4C}{3(E^2-1)^2} - \frac{\alpha_4+12C\alpha_5}{24(r_{04}-2)(r_{03}-2)(r_{02}-2)(r_{02}-r_{04})^2(r_{02}-r_{03})^2(r_{03}-r_{04})^2} \nonumber \\
& \qquad \qquad + \frac{-75E^2+8+149E^4-116E^6+32E^8}{E^2(E^2-1)^4} \bigg] r_{03}(r_{04}-r_{02}) E(k)  + \frac{256r_{04}}{r_{04}-2} \Pi(n_2,k) \nonumber \\
& \quad \quad + \bigg[ \frac{2C(2E^2-5)}{(E^2-1)^2} - \frac{\alpha_6+4C\alpha_7}{8(r_{02}-r_{04})(r_{02}-r_{03})(r_{03}-r_{04})^2(E^2-1)} \nonumber \\
& \qquad \qquad + \frac{238+128E^8+952E^4-754E^2-560E^6}{(E^2-1)^4} \bigg] r_{04} \Pi(n_1,k) \Bigg\}
\end{align}
where $Z=-r_{03}r_{04}K(k)+r_{03}(r_{04}-r_{02})E(k)+2r_{04}\frac{3-2E^2}{1-E^2} \Pi(n_1,k) + \frac{8r_{04}}{r_{04}-2} \Pi(n_2,k)$ and $\gamma_1, \alpha_1,\ldots,\alpha_7$ are non-symmetric expressions of $r_{02}$, $r_{03}$, and $r_{04}$. These constants read
\begin{align}
\gamma_1 = -2r_{03}^2r_{04}^2+3r_{03}^3r_{04}+3r_{04}^3r_{03}-3r_{03}^3r_{02}-3r_{04}^3r_{02}+r_{03}^2r_{02}^2-2r_{03}r_{04}r_{02}^2+r_{04}^2r_{03}r_{02}+r_{04}^2r_{02}^2+r_{03}^2r_{04}r_{02}
\end{align}
\begin{align}
\alpha_1 = & 1536-288r_{04}-288r_{03}+96r_{02}+20r_{02}^2-8r_{04}^2r_{02}+4r_{03}^2r_{04}r_{02}+4r_{04}^2r_{03}r_{02}+5r_{03}r_{04}r_{02}^2 \nonumber \\
& +8r_{02}r_{03}r_{04}+14r_{03}^2r_{04}^2+15r_{04}^3r_{03}-10r_{03}r_{02}^2-10r_{04}r_{02}^2-30r_{03}^3-8r_{03}^2r_{02}+14r_{04}^2r_{03}+14r_{03}^2r_{04} \nonumber  \\
& +56r_{03}r_{04}-32r_{02}r_{04}-32r_{02}r_{03}-84r_{03}^2+15r_{03}^3r_{04}-30r_{04}^3-84r_{04}^2 \,,
\end{align}
\begin{align}
\alpha_2 = & 864r_{03}r_{04}^2r_{02}^2-r_{03}^4r_{04}r_{02}^3-336r_{03}r_{04}^3r_{02}+384r_{03}^2r_{04}r_{02}-20r_{03}^2r_{04}r_{02}^3+384r_{04}^2r_{03}r_{02}+36r_{03}^3r_{04}r_{02}^3 \nonumber \\
& -52r_{02}^3r_{04}^2r_{03}^2+48r_{03}^4r_{04}r_{02}^2+58r_{02}^4r_{04}r_{03}^2-768r_{03}r_{04}r_{02}^2+58r_{02}^4r_{04}^2r_{03}-12r_{03}r_{02}^3r_{04}^3-96r_{02}^3r_{04}r_{03} \nonumber \\
& +286r_{03}^2r_{02}^2r_{04}^3+592r_{03}^2r_{02}r_{04}^3+124r_{02}^3r_{04}^2r_{03}-812r_{03}r_{02}^2r_{04}^3-474r_{04}^4r_{02}r_{03}^2+200r_{03}r_{04}^5r_{02}+96r_{03}r_{04}^4r_{02}^2 \nonumber \\
& +80r_{03}^5r_{04}r_{02}+318r_{03}^4r_{02}r_{04}^2-416r_{03}^3r_{04}^2r_{02}+32r_{03}^3r_{04}^3r_{02}-290r_{03}^3r_{04}^2r_{02}^2+432r_{03}^3r_{04}r_{02}+23r_{03}^5r_{02}^2r_{04} \nonumber \\
& -7r_{03}^5r_{04}^2r_{02}+308r_{03}r_{02}r_{04}^4-232r_{02}^4r_{04}r_{03}-460r_{03}^4r_{04}r_{02}-576r_{03}^2r_{04}^2r_{02}-288r_{03}^2r_{02}^2r_{04}+148r_{03}^3r_{04}r_{02}^2 \nonumber \\
& +29r_{02}^4r_{04}r_{03}^3-13r_{03}r_{04}^4r_{02}^3-11r_{03}^3r_{04}^2r_{02}^3+416r_{03}^4r_{04}^2+414r_{04}^6r_{02}+137r_{02}r_{03}^2r_{04}^5+102r_{03}^6r_{02}+58r_{02}r_{03}^3r_{04}^4 \nonumber \\
& +29r_{02}^4r_{04}^3r_{03}-58r_{02}^4r_{04}^2r_{03}^2+120r_{03}^3r_{02}^2r_{04}^3-768r_{03}^2r_{04}^2-207r_{03}r_{02}r_{04}^6+48r_{02}^3r_{04}^2+25r_{03}^2r_{02}^3r_{04}^3 \nonumber \\
& -76r_{02}^3r_{04}^3-179r_{02}^2r_{03}^2r_{04}^4+95r_{03}r_{04}^5r_{02}^2-35r_{03}^4r_{02}^2r_{04}^2-98r_{03}^4r_{04}^3r_{02}-51r_{03}^6r_{04}r_{02}-948r_{04}^5r_{02} \nonumber \\
& +524r_{02}^2r_{04}^4-832r_{03}^2r_{04}^4-816r_{04}^4r_{03}+816r_{03}^2r_{04}^3-34r_{03}^5r_{04}^2-10r_{03}^2r_{04}^5-368r_{03}^4r_{04}^3-384r_{04}^3r_{02}-414r_{04}^6r_{03} \nonumber \\
& +44r_{03}^4r_{02}^2+116r_{02}^4r_{03}^2-384r_{03}^3r_{02}+48r_{03}^2r_{02}^3+432r_{03}^4r_{02}+352r_{03}^3r_{04}^4-102r_{03}^6r_{04}-336r_{03}^3r_{04}^2+384r_{04}^3r_{03} \nonumber \\
& -58r_{02}^4r_{04}^3-232r_{03}^3r_{04}^5+207r_{03}^2r_{04}^6-132r_{03}^5r_{02}+132r_{03}^5r_{04}-16r_{03}^5r_{04}^3+51r_{03}^6r_{04}^2+134r_{03}^4r_{04}^4+192r_{02}^2r_{03}^2r_{04}^2 \nonumber \\
& -28r_{02}^3r_{03}^3-58r_{03}^3r_{02}^4-432r_{03}^4r_{04}+384r_{04}^2r_{02}^2-46r_{03}^5r_{02}^2+384r_{03}^2r_{02}^2+384r_{03}^3r_{04}-190r_{04}^5r_{02}^2+816r_{04}^4r_{02} \nonumber \\
& +116r_{02}^4r_{04}^2+296r_{03}^3r_{04}^3+26r_{04}^4r_{02}^3+948r_{03}r_{04}^5+2r_{03}^4r_{02}^3-480r_{04}^3r_{02}^2-96r_{02}^2r_{03}^3 \,,
\end{align}
\begin{align}
\alpha_3 = & 32r_{04}^2r_{02}+2r_{03}r_{04}^2r_{02}^2+14r_{03}r_{04}^3r_{02}-32r_{03}^2r_{04}r_{02}-16r_{03}r_{04}r_{02}^2+4r_{03}^2r_{02}r_{04}^3+3r_{03}r_{02}^2r_{04}^3-4r_{03}^3r_{04}^2r_{02} \nonumber \\
& +14r_{03}^3r_{04}r_{02}-11r_{03}r_{02}r_{04}^4-3r_{03}^4r_{04}r_{02}+2r_{03}^2r_{02}^2r_{04}+3r_{03}^3r_{04}r_{02}^2+3r_{03}^4r_{04}^2+24r_{03}^2r_{04}^2+11r_{03}^2r_{04}^4-22r_{04}^4r_{03}\nonumber \\
& -18r_{03}^2r_{04}^3-44r_{04}^3r_{02}-12r_{03}^3r_{02}+6r_{03}^4r_{02}-2r_{03}^3r_{04}^2+44r_{04}^3r_{03}+32r_{03}^2r_{02}-32r_{04}^2r_{03}-32r_{03}^2r_{04}-4r_{02}^2r_{03}^2r_{04}^2\nonumber \\
& -6r_{03}^4r_{04}+12r_{04}^2r_{02}^2+12r_{03}^2r_{02}^2+12r_{03}^3r_{04}+22r_{04}^4r_{02}-2r_{03}^3r_{04}^3-6r_{04}^3r_{02}^2-6r_{02}^2r_{03}^3 \,,
\end{align}
\begin{align}
\alpha_7 = &  -12r_{02}^3r_{04}r_{03}+9r_{03}^2r_{02}^3+9r_{02}^3r_{04}^2-16r_{03}r_{04}r_{02}^2-r_{03}r_{04}^2r_{02}^2-9r_{03}^2r_{02}^2r_{04}+16r_{03}^2r_{02}^2+2r_{02}^2r_{03}^3+16r_{04}^2r_{02}^2\nonumber \\
& -6r_{04}^3r_{02}^2+6r_{03}r_{04}^3r_{02}-32r_{04}^3r_{02}-3r_{03}^4r_{02}+8r_{04}^2r_{03}r_{02}+2r_{03}^2r_{04}^2r_{02}-24r_{03}^2r_{04}r_{02}-11r_{04}^4r_{02}-2r_{03}^3r_{04}r_{02}\nonumber \\
& +3r_{03}^4r_{04}+5r_{03}^2r_{04}^3-3r_{03}^3r_{04}^2+11r_{04}^4r_{03}+32r_{04}^3r_{03} \,.
\end{align}
\begin{align}
\alpha_4 = &  204r_{03}^7r_{04}^2-128r_{03}^5r_{02}^4+348r_{02}^7r_{04}^2+80r_{03}^6r_{02}^4+32r_{02}^5r_{04}^3+560r_{03}^4r_{04}r_{02}^3-192r_{02}^5r_{04}r_{03}-1536r_{03}^2r_{04}r_{02}^3\nonumber \\
& -1344r_{03}^3r_{04}r_{02}^3+1344r_{02}^3r_{04}^2r_{03}^2+2400r_{03}^4r_{04}r_{02}^2+864r_{02}^4r_{04}r_{03}^2-1440r_{02}^4r_{04}^2r_{03}+1728r_{03}r_{02}^3r_{04}^3\nonumber \\
& -4800r_{03}^2r_{02}^2r_{04}^3-1536r_{03}^2r_{02}r_{04}^3-1536r_{02}^3r_{04}^2r_{03}-1536r_{03}r_{02}^2r_{04}^3+864r_{04}^4r_{02}r_{03}^2-3264r_{03}r_{04}^5r_{02}\nonumber \\
& +3168r_{03}r_{04}^4r_{02}^2-1728r_{03}^5r_{04}r_{02}-2208r_{03}^4r_{02}r_{04}^2-1536r_{03}^3r_{04}^2r_{02}+1728r_{03}^3r_{04}^3r_{02}+1344r_{03}^3r_{04}^2r_{02}^2\nonumber \\
& -1088r_{03}^5r_{02}^2r_{04}+1792r_{03}^5r_{04}^2r_{02}+1536r_{03}r_{02}r_{04}^4+1536r_{02}^4r_{04}r_{03}+1536r_{03}^4r_{04}r_{02}-264r_{03}^6r_{02}^2+68r_{03}^6r_{04}^3\nonumber \\
& +828r_{03}^2r_{04}^7-704r_{03}^4r_{04}^5+128r_{03}^3r_{02}^5-414r_{03}^3r_{04}^7+464r_{03}^4r_{04}^6-1536r_{03}^3r_{04}r_{02}^2-112r_{02}^4r_{04}r_{03}^3-220r_{04}^6r_{02}^3\nonumber \\
& +20r_{03}^3r_{04}^6-496r_{02}^4r_{04}^4-2032r_{03}r_{04}^4r_{02}^3+1424r_{03}^3r_{04}^2r_{02}^3+96r_{03}^2r_{02}^5-768r_{03}^4r_{04}^2-800r_{02}r_{03}^2r_{04}^5-1840r_{02}r_{03}^3r_{04}^4\nonumber \\
& +1232r_{02}^4r_{04}^3r_{03}-608r_{02}^4r_{04}^2r_{03}^2-1264r_{03}^3r_{02}^2r_{04}^3+3792r_{03}r_{02}r_{04}^6-1264r_{03}^2r_{02}^3r_{04}^3+1536r_{02}^3r_{04}^3+4960r_{02}^2r_{03}^2r_{04}^4\nonumber \\
& -3392r_{03}r_{04}^5r_{02}^2-2240r_{03}^4r_{02}^2r_{04}^2+2096r_{03}^4r_{04}^3r_{02}+528r_{03}^6r_{04}r_{02}-768r_{02}^2r_{04}^4-768r_{03}^2r_{04}^4+864r_{03}^5r_{04}^2+1632r_{03}^2r_{04}^5\nonumber \\
& +672r_{03}^4r_{04}^3+96r_{02}^5r_{04}^2-768r_{03}^4r_{02}^2-768r_{02}^4r_{03}^2-1632r_{03}^3r_{04}^4-292r_{03}r_{04}^6r_{02}^4+266r_{03}^3r_{04}^4r_{02}^4-190r_{03}^2r_{04}^4r_{02}^5\nonumber \\
& +122r_{03}^2r_{04}^5r_{02}^4+170r_{03}r_{04}^5r_{02}^5+266r_{02}^3r_{03}^4r_{04}^4+64r_{03}^6r_{02}^3r_{04}^2-40r_{03}^6r_{04}r_{02}^4-620r_{02}^3r_{03}^3r_{04}^5+207r_{03}r_{04}^7r_{02}^3\nonumber \\
& +352r_{03}^2r_{04}^6r_{02}^3-58r_{03}^5r_{02}^4r_{04}^2-114r_{02}^6r_{04}^3r_{03}+87r_{03}^3r_{04}r_{02}^7-68r_{03}^2r_{04}r_{02}^6+172r_{03}^3r_{04}^6r_{02}^2-414r_{03}^2r_{04}^7r_{02}^2\nonumber \\
& +756r_{03}^4r_{04}^2r_{02}^4+102r_{03}^7r_{04}r_{02}^2+464r_{03}^5r_{04}r_{02}^3+1328r_{03}r_{04}^5r_{02}^3+460r_{03}r_{04}^6r_{02}^2+38r_{02}^5r_{03}^5r_{04}-16r_{02}r_{03}^6r_{04}^4\nonumber \\
& -232r_{03}^4r_{04}^6r_{02}-76r_{03}r_{04}^4r_{02}^6-1656r_{03}r_{04}^7r_{02}+216r_{03}r_{04}^4r_{02}^4-808r_{02}^3r_{03}^2r_{04}^4-384r_{02}^2r_{03}^2r_{04}^5+828r_{04}^7r_{02}^2\nonumber \\
& -164r_{03}^6r_{02}r_{04}^2+64r_{02}^5r_{04}^2r_{03}-8r_{03}^6r_{04}^3r_{02}^2-114r_{03}^6r_{04}r_{02}^3+176r_{03}^4r_{02}^4+192r_{03}^5r_{02}^2r_{04}^2-18r_{03}^6r_{04}^3r_{02}\nonumber \\
& +20r_{03}^6r_{02}^2r_{04}^2-148r_{02}^6r_{04}^2r_{03}^2-544r_{03}^4r_{02}^3r_{04}^3-354r_{03}^3r_{04}^6r_{02}-1096r_{02}^2r_{03}^3r_{04}^4-784r_{02}^5r_{04}^3r_{03}-260r_{03}^2r_{04}^6r_{02}\nonumber \\
& -174r_{02}^7r_{04}^2r_{03}^2-1896r_{02}^2r_{04}^6-262r_{03}^4r_{02}^4r_{04}^3+74r_{03}^4r_{04}^2r_{02}^5+102r_{03}^7r_{02}r_{04}^2+87r_{02}^7r_{04}^3r_{03}+51r_{03}^7r_{04}^3r_{02}\nonumber \\
& -102r_{03}^7r_{04}^2r_{02}^2+174r_{02}^7r_{04}^2r_{03}+194r_{03}^4r_{02}^2r_{04}^5-20r_{02}^5r_{04}^3r_{03}^3+112r_{02}^6r_{04}^3r_{03}^2+414r_{03}r_{04}^7r_{02}^2+64r_{03}^4r_{02}^5\nonumber \\
& -594r_{03}r_{04}^6r_{02}^3-516r_{03}^2r_{04}^4r_{02}^4+252r_{03}r_{04}^5r_{02}^4+414r_{03}^2r_{04}^7r_{02}-174r_{03}^3r_{02}^7-102r_{03}^7r_{04}^3-28r_{03}^6r_{02}^3+132r_{02}r_{03}^5r_{04}^4\nonumber \\
& -176r_{02}^5r_{04}^2r_{03}^3+568r_{03}^5r_{04}^3r_{02}^2+174r_{03}^2r_{04}r_{02}^7+336r_{02}^6r_{04}r_{03}-536r_{03}^5r_{02}^3r_{04}^2+424r_{02}^5r_{04}^3r_{03}^2+60r_{03}r_{04}^4r_{02}^5\nonumber \\
& +180r_{03}^5r_{04}r_{02}^4-552r_{03}^4r_{04}r_{02}^4-76r_{02}^6r_{03}^3-408r_{03}^7r_{04}r_{02}+51r_{03}^7r_{04}r_{02}^3+207r_{03}^3r_{04}^7r_{02}+30r_{03}^3r_{04}r_{02}^6\nonumber \\
& -250r_{02}^2r_{03}^5r_{04}^4+4r_{02}^6r_{04}^2r_{03}^3+136r_{03}^5r_{04}^3r_{02}^3-180r_{03}^4r_{04}r_{02}^5-40r_{03}^4r_{04}r_{02}^6+448r_{03}^3r_{02}^2r_{04}^5+134r_{03}^5r_{04}^5r_{02}\nonumber \\
& +364r_{02}^6r_{04}^2r_{03}-224r_{03}^2r_{04}r_{02}^5+368r_{03}^3r_{04}r_{02}^5+124r_{03}^6r_{04}r_{02}^2-520r_{02}^4r_{04}^2r_{03}^3-232r_{02}^4r_{04}^3r_{03}^3-36r_{02}r_{03}^4r_{04}^5\nonumber \\
& -540r_{02}^2r_{03}^4r_{04}^4-1264r_{03}^5r_{04}^3r_{02}+312r_{02}r_{03}^4r_{04}^4+752r_{02}r_{03}^3r_{04}^5+920r_{03}^4r_{02}^2r_{04}^3+728r_{02}^4r_{04}^3r_{03}^2+552r_{03}^3r_{02}^3r_{04}^3\nonumber \\
& -40r_{03}^4r_{04}^2r_{02}^3-100r_{02}^2r_{03}^2r_{04}^6+680r_{02}^3r_{03}^3r_{04}^4-56r_{02}^3r_{03}^2r_{04}^5-696r_{02}^7r_{04}r_{03}+2528r_{04}^5r_{02}^3-76r_{02}^5r_{03}^5-992r_{04}^5r_{02}^4\nonumber \\
& -168r_{03}^2r_{02}^6-168r_{02}^6r_{04}^2+32r_{03}^6r_{04}^4+204r_{03}^7r_{02}^2+640r_{02}^5r_{04}^4-220r_{02}^6r_{04}^3-340r_{04}^5r_{02}^5+584r_{04}^6r_{02}^4-414r_{04}^7r_{02}^3\nonumber \\
& -174r_{02}^7r_{04}^3+152r_{04}^4r_{02}^6+128r_{03}^5r_{02}^3+348r_{03}^2r_{02}^7+672r_{02}^4r_{04}^3+1664r_{03}^3r_{04}^5-1896r_{03}^2r_{04}^6-268r_{03}^5r_{04}^5-102r_{03}^7r_{02}^3\nonumber \\
& +80r_{03}^4r_{02}^6+736r_{03}^5r_{04}^4-832r_{03}^5r_{04}^3-264r_{03}^6r_{04}^2-592r_{03}^4r_{04}^4+4608r_{02}^2r_{03}^2r_{04}^2+1536r_{02}^3r_{03}^3-96r_{03}^3r_{02}^4\nonumber \\
& +864r_{03}^5r_{02}^2+1632r_{04}^5r_{02}^2-768r_{02}^4r_{04}^2+1536r_{03}^3r_{04}^3-2400r_{04}^4r_{02}^3-864r_{03}^4r_{02}^3 \,,
\end{align}
\begin{align}
\alpha_5 = & -18r_{02}^5r_{04}^3+128r_{03}r_{04}^2r_{02}^2-16r_{03}^4r_{04}r_{02}^3-128r_{03}r_{04}^3r_{02}-48r_{02}^5r_{04}r_{03}+24r_{03}^2r_{04}r_{02}^3+24r_{03}^3r_{04}r_{02}^3+32r_{02}^3r_{04}^2r_{03}^2\nonumber \\
& +24r_{03}^4r_{04}r_{02}^2-40r_{02}^4r_{04}r_{03}^2-8r_{02}^4r_{04}^2r_{03}+56r_{03}r_{02}^3r_{04}^3+16r_{03}^2r_{02}^2r_{04}^3+120r_{03}^2r_{02}r_{04}^3-40r_{02}^3r_{04}^2r_{03}\nonumber \\
& -168r_{03}r_{02}^2r_{04}^3-56r_{04}^4r_{02}r_{03}^2-88r_{03}r_{04}^5r_{02}+24r_{03}r_{04}^4r_{02}^2-24r_{03}^5r_{04}r_{02}-24r_{03}^4r_{02}r_{04}^2+120r_{03}^3r_{04}^2r_{02}\nonumber \\
& -32r_{03}^3r_{04}^3r_{02}-16r_{03}^3r_{04}^2r_{02}^2-128r_{03}^3r_{04}r_{02}+6r_{03}^5r_{02}^2r_{04}+6r_{03}^5r_{04}^2r_{02}+176r_{03}r_{02}r_{04}^4+32r_{02}^4r_{04}r_{03}+48r_{03}^4r_{04}r_{02}\nonumber \\
& -128r_{03}^2r_{04}^2r_{02}+128r_{03}^2r_{02}^2r_{04}-18r_{03}^3r_{02}^5-104r_{03}^3r_{04}r_{02}^2+20r_{02}^4r_{04}r_{03}^3+24r_{02}^4r_{04}^4-32r_{03}r_{04}^4r_{02}^3-16r_{03}^3r_{04}^2r_{02}^3\nonumber \\
& +36r_{03}^2r_{02}^5-24r_{03}^4r_{04}^2+22r_{02}r_{03}^2r_{04}^5-12r_{02}r_{03}^3r_{04}^4+20r_{02}^4r_{04}^3r_{03}+20r_{02}^4r_{04}^2r_{03}^2+32r_{03}^3r_{02}^2r_{04}^3-64r_{02}^3r_{04}^2\nonumber \\
& -48r_{03}^2r_{02}^3r_{04}^3+96r_{02}^3r_{04}^3+16r_{02}^2r_{03}^2r_{04}^4+22r_{03}r_{04}^5r_{02}^2+4r_{03}^4r_{04}^3r_{02}-88r_{02}^2r_{04}^4-88r_{03}^2r_{04}^4+64r_{03}^2r_{04}^3+12r_{03}^5r_{04}^2\nonumber \\
& +44r_{03}^2r_{04}^5+4r_{03}^4r_{04}^3+36r_{02}^5r_{04}^2-24r_{03}^4r_{02}^2-8r_{02}^4r_{03}^2-64r_{03}^2r_{02}^3+36r_{03}^3r_{04}^4+64r_{03}^3r_{04}^2+3r_{03}^5r_{04}r_{02}^3\nonumber \\
& +11r_{03}r_{04}^5r_{02}^3+6r_{02}^5r_{04}^2r_{03}+8r_{03}^4r_{02}^4-6r_{03}^5r_{02}^2r_{04}^2-3r_{02}^2r_{03}^3r_{04}^4+9r_{02}^5r_{04}^3r_{03}-12r_{02}^5r_{04}^2r_{03}^2-4r_{03}^4r_{04}r_{02}^4\nonumber \\
& +6r_{03}^2r_{04}r_{02}^5+9r_{03}^3r_{04}r_{02}^5-7r_{02}^4r_{04}^2r_{03}^3+3r_{03}^5r_{04}^3r_{02}-2r_{02}r_{03}^4r_{04}^4+11r_{02}r_{03}^3r_{04}^5-3r_{03}^4r_{02}^2r_{04}^3+r_{02}^4r_{04}^3r_{03}^2\nonumber \\
& -2r_{03}^3r_{02}^3r_{04}^3+9r_{03}^4r_{04}^2r_{02}^3-12r_{03}r_{04}^4r_{02}^4+17r_{02}^3r_{03}^2r_{04}^4-22r_{02}^2r_{03}^2r_{04}^5-22r_{04}^5r_{02}^3-6r_{03}^5r_{02}^3-44r_{02}^4r_{04}^3-22r_{03}^3r_{04}^5\nonumber \\
& -6r_{03}^5r_{04}^3+4r_{03}^4r_{04}^4-48r_{02}^2r_{03}^2r_{04}^2+32r_{02}^3r_{03}^3-12r_{03}^3r_{02}^4+12r_{03}^5r_{02}^2+44r_{04}^5r_{02}^2-8r_{02}^4r_{04}^2-48r_{03}^3r_{04}^3-4r_{04}^4r_{02}^3\nonumber\\
& -4r_{03}^4r_{02}^3+64r_{04}^3r_{02}^2+64r_{02}^2r_{03}^3 \,,
\end{align}
\begin{align}
\alpha_6 = & 32r_{03}r_{04}^3r_{02}-58r_{02}^5r_{04}r_{03}+128r_{03}^2r_{04}r_{02}-28r_{03}^2r_{04}r_{02}^3+128r_{04}^2r_{03}r_{02}+5r_{03}^3r_{04}r_{02}^3+26r_{02}^3r_{04}^2r_{03}^2\nonumber \\
& -35r_{03}^4r_{04}r_{02}^2-18r_{02}^4r_{04}r_{03}^2-256r_{03}r_{04}r_{02}^2+18r_{02}^4r_{04}^2r_{03}-67r_{03}r_{02}^3r_{04}^3-64r_{02}^3r_{04}r_{03}+r_{03}^2r_{02}^2r_{04}^3\nonumber \\
& +128r_{03}^2r_{02}r_{04}^3+68r_{02}^3r_{04}^2r_{03}-260r_{03}r_{02}^2r_{04}^3+5r_{04}^4r_{02}r_{03}^2+49r_{03}r_{04}^5r_{02}+5r_{03}r_{04}^4r_{02}^2+9r_{03}^5r_{04}r_{02}\nonumber \\
& -39r_{03}^4r_{02}r_{04}^2-160r_{03}^3r_{04}^2r_{02}+30r_{03}^3r_{04}^3r_{02}+5r_{03}^3r_{04}^2r_{02}^2+32r_{03}^3r_{04}r_{02}+140r_{03}r_{02}r_{04}^4-88r_{02}^4r_{04}r_{03}\nonumber \\
& -52r_{03}^4r_{04}r_{02}-4r_{03}^3r_{04}r_{02}^2+29r_{03}^2r_{02}^5+16r_{03}^4r_{04}^2-69r_{04}^6r_{02}-17r_{03}^6r_{02}-256r_{03}^2r_{04}^2+32r_{02}^3r_{04}^2-36r_{02}^3r_{04}^3\nonumber \\
& -236r_{04}^5r_{02}+164r_{02}^2r_{04}^4-304r_{03}^2r_{04}^4+32r_{04}^4r_{03}-32r_{03}^2r_{04}^3-11r_{03}^5r_{04}^2-15r_{03}^2r_{04}^5+62r_{03}^4r_{04}^3-128r_{04}^3r_{02}\nonumber \\
& +69r_{04}^6r_{03}+29r_{02}^5r_{04}^2+36r_{03}^4r_{02}^2+44r_{02}^4r_{03}^2-128r_{03}^3r_{02}+32r_{03}^2r_{02}^3-32r_{03}^4r_{02}-58r_{03}^3r_{04}^4+17r_{03}^6r_{04}\nonumber \\
& -32r_{03}^3r_{04}^2+128r_{04}^3r_{03}-6r_{02}^4r_{04}^3-12r_{03}^5r_{02}+12r_{03}^5r_{04}+128r_{02}^2r_{03}^2r_{04}^2-4r_{02}^3r_{03}^3+6r_{03}^3r_{02}^4+32r_{03}^4r_{04}\nonumber \\
& +128r_{04}^2r_{02}^2+2r_{03}^5r_{02}^2+128r_{03}^2r_{02}^2+128r_{03}^3r_{04}-34r_{04}^5r_{02}^2-32r_{04}^4r_{02}+44r_{02}^4r_{04}^2+168r_{03}^3r_{04}^3\nonumber \\
& +48r_{04}^4r_{02}^3+236r_{03}r_{04}^5+12r_{03}^4r_{02}^3 \,,
\end{align}
\end{widetext}
These complicated expressions are exact and may be simplified by an additional approximation. For example, for large peri- and apoapsis only the higher orders of $r_{03}$ and $r_{04}$ may be taken into account. In particular, in the post-Newtonian case these long expressions are reduced to \eqref{OmegaPLambda_PN}.

\bibliographystyle{unsrt}
\bibliography{PD_obs}

\end{document}